\documentclass[twocolumn]{pasj00}
\usepackage[varg]{txfonts}
\color{black}

\newcommand{\fluxunit}{erg~cm$^{-2}$~s$^{-1}$}
\newcommand{\sbunit}{erg~cm$^{-2}$~s$^{-1}$~sr$^{-1}$}

\begin{document}
\Received{$\langle$reception date$\rangle$}
\Accepted{$\langle$acception date$\rangle$}
\Published{$\langle$publication date$\rangle$}
\SetRunningHead{H. Akamatsu et al.}{Properties of intracluster medium of  Abell 3667 Observed with Suzaku XIS}
\title{
Properties of the intracluster medium of   Abell 3667 observed with Suzaku XIS\thanks{Last update: \today}}
\author{
 H. Akamatsu\altaffilmark{1},
 J. de Plaa\altaffilmark{2},
 J. Kaastra\altaffilmark{2},
 Y. Ishisaki\altaffilmark{1},
 T. Ohashi\altaffilmark{1},
 M. Kawaharada\altaffilmark{3},
 K. Nakazawa\altaffilmark{4}, 
}
\altaffiltext{1}{
Department of Physics, Tokyo Metropolitan University,\\
 1-1 Minami-Osawa, Hachioji, Tokyo 192-0397}
\altaffiltext{2}{
SRON Netherlands Institute for Space Research, Sorbonnelaan 2, 3584 CA Utrecht, The Netherlands
}
 \altaffiltext{3}{
Department of High Energy Astrophysics, Institute of Space and 
Astronautical Science,\\ Japan Aerospace Exploration Agency,\\
 3-1-1 Yoshinodai,Sagamihara, Kanagawa 229-8510}
\email{h\_aka@phys.metro-u.ac.jp}
\KeyWords{
galaxies: clusters: individual (Abell 3667)
--- X-rays:ICM, shock wave}
\maketitle
\begin{abstract}
We observed the northwest region of the cluster of galaxies A3667 with the Suzaku XIS instrument.  
The temperature and surface brightness of the intracluster medium were measured up to the virial radius
 ($r_{200}\sim 2.3$ Mpc).  
The radial temperature profile is flatter than the average profile for other clusters until the radius reaches the northwest radio relic.  
The temperature drops sharply from 5 keV to about 2 keV at the northwest radio relic region.  
The sharp changes of the temperature can be interpreted as a shock with a Mach number of about 2.5.  
The entropy slope becomes flatter in the outer region and negative around the radio relic region.  
In this region, the relaxation timescale of electron-ion Coulomb collisions is longer than the time elapsed after the shock heating and 
the plasma may be out of equilibrium.
Using differential emission measure (DEM) models, we also confirm the multi-temperature structure around the radio relic region, characterized by two peaks at 1 keV and 4 keV\@.  
These features suggest that the gas is heated by a shock propagating from the center to outer region.
\end{abstract}

\section{Introduction}
\label{sec:intro}
Clusters of galaxies grow through gravitational infall and mergers of smaller groups and clusters.  
Merger events convert kinetic and turbulent energy of the gas in colliding subclusters into thermal energy by driving shocks in the cluster.  A fundamental issue of cluster growth is the nature of heating processes in merger events.  
Although shocks caused by merger events are very important for gas heating and particle acceleration, there are only a few clusters for
which clear evidence of shocks has been obtained (1E0657-56, A520:~\cite{clowe06,markevitch05}).

An additional science issue of merging clusters is the temperature structure.
In dynamically young systems, the intracluster medium (ICM) is thought to be in an early stage of thermal relaxation.  
Some theoretical studies predict non-ionization equilibrium states and an electron-ion two-temperature structure 
in the ICM of merging galaxy clusters~\citep{rudd09,akahori10}.  
Recent Suzaku results show that the temperature profile of the merging cluster A2142 is slightly in
excess compared to other relaxed clusters and a possible deviation between electron and 
ion temperatures may be present~\citep{akamatsu11}.  
However, there is little information about the ICM properties in the outer regions of merging clusters because of the faintness of the X-ray
surface brightness.

In this paper, we present results from observations of the merging cluster A3667 ($z = 0.0556$) with XIS on board Suzaku
\citep{mitsuda07}.  
It is a very bright merging cluster with irregular morphology and an average temperature of $7.0 \pm 0.5$ keV \citep{knopp96}.  
\citet{briel04} studied the ICM characteristics in the cluster central region and showed that 
the ICM emission is elongated to the northwest direction.  
The most striking features of A3667 is a discontinuity in the X-ray surface brightness, namely a
``cold front'', which was found with Chandra \citep{vikhlinin01}.
Another feature of A3667 are the two extended, symmetrically
located regions of diffuse radio emission revealed with 843 MHz ATCA observatory 
(Australia Telescope Compact Array: \cite{rottgering97}). 
The presence of these structures naturally indicates that A3667 is a merging system
and the subcluster infall seems to be occurring along the northwest--southeast direction.  
Actually, the overall cluster emission is elongated in this direction.

\begin{table*}[t]
\label{tab:obslog}
\begin{center}
\caption{Suzaku observation log of Abell 3667}
\begin{tabular}{cccccccccc}
\hline
Region(Obs.ID) & Obs. date &Raw Exposure$^\ast$ &Cleaned Exposure$^\dagger$ \\ 
			&			&(ks)			&(ks) \\
\hline
CENTER  (801055010)  	& 2006-5-6T17:40:17	&19.3	&16.8\\
OFFSET1 (802030010)  	& 2006-5-6T07:03:14	&16.2	&11.8\\
OFFSET2 (802031010)  	& 2006-5-3T17:47:01	&81.3	&60.3\\
\hline
$\ast$:COR2  $>$ 0 GV &\multicolumn{2}{l}{ ${\dagger}$:COR2 $>$ 8 GV and $\rm ELV\>10^{\circ}$}\\\
\end{tabular}
\end{center}
\end{table*}

A recent XMM-Newton observation of the northwest radio relic of A3667 has shown
significant jumps in temperature and surface brightness across the
relic~\citep{finoguenov10}.  The observed ICM properties associated
with the jump indicate the existence of a shock front with a Mach number ${\cal M} \sim 2.4$.

From Suzaku XIS/HXD observations of A3667, \citet{nakazawa09}
measured the cluster emission up to \timeform{40'} and set upper
limits on the non-thermal X-ray emission in XIS spectra of $7.3\times 10^{-13}$~\fluxunit~when extrapolated to 10-40 keV\@.
\citet{nakazawa09} also observed very high temperature components
($kT\sim 14$ keV) in the cluster center.  Since their main interest
was to look into the non-thermal phenomenon, the global ICM properties
in the cluster outskirts remain to be further explored.

The purpose of this study is to clarify the physical state of the ICM in
the outer regions of the merging cluster A3667.  
We use $H_0 = 70$ km s$^{-1}$ Mpc$^{-1}$, $\Omega_{\rm M}=0.27$ and
$\Omega_\Lambda = 0.73$, respectively, which means that 1' correspond to a diameter of 66 kpc.
The virial radius is approximated by $r_{200} = 2.77 h_{70}^{-1} (\langle T\rangle /10 {\rm keV})^{1/2}{\rm Mpc} /E(z)$,
where $E(z)=(\Omega_{\rm M}(1+z)^{3}+1-\Omega_{\rm M})^{1/2}$ \citep{henry09}.  
For our cosmology and redshift, $r_{200}$ is 2.26 Mpc ($= \timeform{34.1'}$) with $kT = 7.0$ keV\@.  
In this paper, we employ solar abundances given by \citet{anders89} and 
Galactic hydrogen column  density of $N_{\rm H}=4.7 \times 10^{20}$~cm$^{-2}$~\citep{dickey90}.  
Unless otherwise stated, the errors correspond to 90\% confidence for a single parameter.

\section{Observations \& Data reduction}

\begin{figure}[t]
\begin{center}
\includegraphics[scale=0.3]{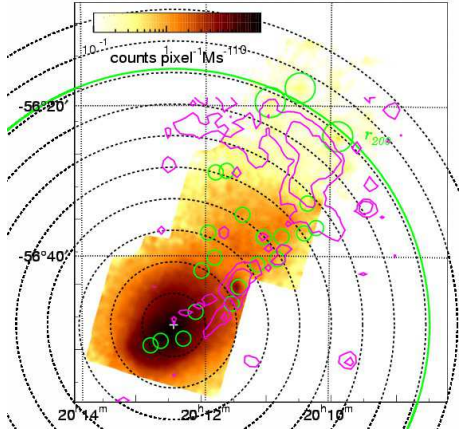}
\end{center}
\caption{
X-ray image of the northwest region of A3667 in the energy band 0.5-8.0 keV, after subtraction of the NXB with no vignetting
 correction and smoothing by a 2-dimensional gaussian with $\sigma =16$ pixel =$\timeform{17''}$. 
 The large green circle indicates the virial radius of A3667, and black dotted circles show 
 the annular regions used for the spectral analysis. 
 Small green circles show point sources that are detected with XMM-Newton. 
 The SUMMS 842 MHz radio image is shown with magenta contours.
}
\label{fig:suzaku_image}
\end{figure}
As shown in Fig~\ref{fig:suzaku_image}, Suzaku carried out three pointing observations of Abell 3667 along the northwest merger
axis in May 2006, designated as Center, Offset1, and Offset2.  
The observation log is summarized in Table~\ref{tab:obslog}.
All observations were performed with either normal $5\times5$ or $3\times3$ clocking mode.

The combined observed field extends to the virial radius of A3667 ($34.1^{'} \sim 2.26$ Mpc).  
During the observations, the contamination from the Solar Wind Charge eXchange (SWCX) was not significant, 
which we base on ACE SWE data and the absence of flare-like features in the XIS light curve (see Appendix 1).

The XIS instrument consists of 4 CCD chips: one back-illuminated (BI: XIS1) 
and three front-illuminated (FI: XIS0, XIS2, XIS3) ones.  
The IR/UV blocking filters had a significant contamination accumulated at
the time of the observations, and we included its effect and uncertainty on the soft X-ray effective area in our analysis.  
We used HEAsoft version 6.9 and CALDB 2010-12-06 for all the Suzaku data analysis presented here.
In the XIS data analysis, we limited the Earth rim ELEVATION $> 10^{\circ}$ to 
avoid contamination of scattered solar X-rays from the day Earth limb.  
Additionally, we performed event screening with cut-off rigidity (COR) $> 8$ GV to increase the signal to noise ratio.
We extracted pulse-height spectra in 10 annular regions whose boundary
radii were $~\timeform{4.2'}, ~\timeform{8.4'}, ~\timeform{12.6'},
~\timeform{16.8'}, ~\timeform{21.0'}, ~\timeform{25.2'},
~\timeform{29.4'}, ~\timeform{33.6'},
\timeform{37.8'} \rm ~and~ \timeform{42.0'}$, with the center at ($\timeform{20h12m31s}, \timeform{-56D49m12s}$).  
We analyzed the spectra in the 0.5--10~keV range for the FI detectors and 0.5--8~keV
for the BI detector.  In all annuli, positions of the calibration sources were masked out using the {\it calmask} calibration database (CALDB) file.

\begin{figure}[t]
\begin{center}
\includegraphics[scale=0.33,angle=-90]{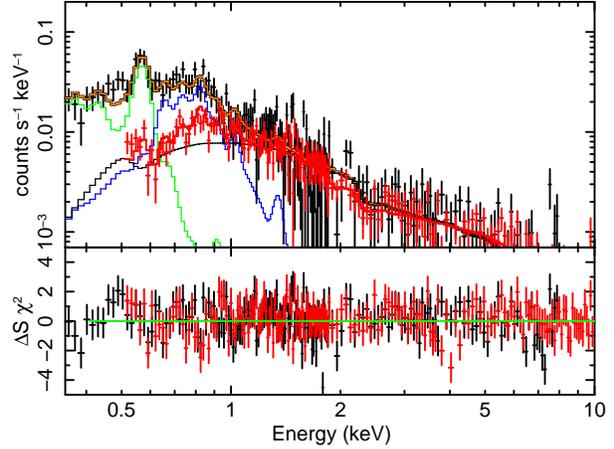}
\end{center}
\caption{ 
The spectrum of the outermost region used for the background
 estimation ($r=\timeform{37.8'}-\timeform{42.0'}$), after NXB
 subtraction.  The XIS BI (Black) and FI (Red) spectra are fitted
 with CXB + Galactic components (LHB, MWH) ({\it apec+wabs(apec+powerlow)}).  The CXB spectrum is shown with a
 black curve, and the LHB and MWH components are indicated by green and blue curves, respectively. 
 }
\label{fig:bgd}
\end{figure}

\begin{table*}[t]
\begin{center}
\small
\caption{Estimation of the CXB fluctuation.}
\begin{tabular}{cccccccccccccc}\hline
 Region   & \timeform{0'}-\timeform{4.2'} & \timeform{4.2'}-\timeform{8.4'} &\timeform{8.4'}-\timeform{12.6'} &\timeform{12.6'}-\timeform{16.8'} & \timeform{16.8'}-\timeform{21.0'} & \timeform{21.0'}-\timeform{25.2'} & \timeform{25.2'}-\timeform{29.4'} & \timeform{29.4'}-\timeform{33.6'}  &   \timeform{33.6'}-\timeform{37.8'}\\ \hline
 $SRR^{\ast}$	&22.28 	&20.89 	&3.53 	&0.42 	&0.29 	&0.16 	&0.05 	&0.04 	&0.03\\  
$\Omega_{\rm e,Suzaku}^{\dagger}$	&50.3 	&158.9 	&73.2 	&67.7 	&80.3 	&72.1 	&59.9 	&62.5 	&60.0\\
${\sigma/I_{\rm CXB}}\ddagger$ 	&14.0 	&7.9 	&11.6 	&12.1 	&11.1 	&11.7 	&12.8 	&12.6 	&12.8 \\
\hline
\multicolumn{10}{l}{$\ast:SROUS\_RATIO\_REG$[\%]:The fraction of simulated cluster photons that full in the region compared to the total photons
}\\
\multicolumn{10}{l}{
 generated in the entire simulation.}\\
\multicolumn{1}{l}{$\dagger$:[arcmin$^{2}$]} \\
\multicolumn{10}{l}{$\ddagger$:[\%]:   In this paper, $S_{c}:=5 \times10^{-14}$ erg cm$^{-2}$ s$^{-1}$is assumed for all regions. $S_{c}$ is the flux limit of the extracted point sources.}\\ 
\label{tab:cxb_fluc}
\end{tabular}
\end{center}
\end{table*}

\begin{table*}[t]
\caption{ Best-fit background parameters}
\begin{center}
\begin{tabular}{lcccccccccccc}
\hline \hline
	& NOMINAL & CXBMAX &CXBMIN &CONTAMI+10\% &CONTAMI-10\% \\ \hline
 LHB& \\
  $kT(\rm keV)$	&	$0.11\pm0.01$	& $0.11\pm0.01$ &$0.10\pm0.01$ & $0.11\pm0.01$	&$0.10\pm0.01$\\
$norm^{\ast}$ ($\times 10^{-3}$)	&	4.85$_{-1.01}^{+1.70}$ &4.48$_{-1.02}^{+1.51}$ & 5.59$_{-1.80}^{+1.54}$ &	4.25$_{-1.20}^{+1.30}$& 6.24$_{-2.10}^{+1.61}$\\ \hline
  MWH &			\\
    $kT(\rm keV)$	&	$0.37\pm0.03$	&$0.37\pm0.04$&$0.37\pm0.04$&$0.37\pm0.04$&$0.36\pm0.04$	\\
$norm^{\ast}$  ($\times 10^{-4}$)&	$11.6_{-1.6}^{+1.2}$	&$11.3_{-1.7}^{+1.1}$&$11.3_{-1.8}^{+1.2}$&	$11.0_{-1.8}^{+1.1}$	&	$12.2_{-2.0}^{+1.2}$		\\ \hline
$\chi^{2}$/d.o.f	& 381/344 & 398/344&406/344&381/344 &382/344 \\ \hline
\multicolumn{6}{l}{\footnotesize
*:Normalization of the apec component scaled with a factor 1/400$\pi$.}\\
\multicolumn{6}{l}{\footnotesize
Norm=$\rm \frac{1}{400\pi}$$\int n_{e}n_{H} dV/(4\pi(1+z^2)D_{A}^2)\times 10 ^{-16} \rm cm^{-5}arcmin^{-2}$, where $D_A$ is the angular diameter distance to the source.}\\
\end{tabular}
\label{tab:bgd}
\end{center}
\end{table*}
\section{Background modeling}
An accurate measurement of the background components is important for the
ICM study in the cluster outer region.  
We examined four background components, namely non-X-ray background (NXB), 
cosmic X-ray background (CXB) and the Galactic emission consisting of the Milky Way Halo (MWH) 
and the Local Hot Bubble (LHB)\@.  
The NXB component was estimated from the dark Earth database by the ${\it xisnxbgen}$ FTOOLS \citep{tawa08} 
and was subtracted from the data before the spectral fit. 
To adjust for the long-term variation of the XIS background due to radiation damage,
we accumulated the NXB data for the period between 150 days before till 150 days after the observation of A3667.

\subsection{Cosmic X-ray Background Intensity and Fluctuation}\label{sec:CXB}
Using seven XMM-Newton data sets reported in \citet{briel04} and \citet{finoguenov10}, 
we searched for point-like sources within the  Suzaku field of view using
the $wavdetect$ tool in the CIAO package.  
As shown by the green circles in figure~\ref{fig:suzaku_image}, 
we detected and subtracted 22 point sources which have 2.0-10 keV fluxes higher than $S_c = 5\times~10^{-14}$~\fluxunit.  
The extraction radius was set to $1' $ or $2'$ corresponding to the HPD of the Suzaku XRT\@.  
We estimated the CXB surface brightness after the source subtraction to be $5.97 \times 10^{-8} $
\sbunit\ based on ASCA GIS measurements \citep{kushino02}.

\citet{kushino02} set the flux-limit of eliminated point sources to be
$S_c=2\times10^{-13}$ erg cm$^{-2}$ s$^{-1}$, 
we subtracted sources down to $5 \times10^{-14}$ erg cm$^{-2}$ s$^{-1}$. 
This causes our flux-limit value to be sufficiently lower than the Kushino et al.\ level.

We need to estimate the amplitude of the CXB fluctuation, 
caused by the statistical fluctuation of point sources in the FOV\@. 
We scaled the measured fluctuations from Ginga (\cite{hayashida89}) to our flux
limit and field of view following \citet{hoshino10}.  The fluctuation
width scales as $-0.5$ power of the effective field of view
($\Omega_{\rm e}$) and $0.25$ power of the point-source detection limit.
We show the CXB fluctuation level, $\Omega_{\rm e}$ 
and $\sigma$/$I_{\rm CXB}$, for each spatial region in table~\ref{tab:cxb_fluc}.

\subsection{Galactic Components}\label{sec:GAL}
To estimate the Galactic background components, we use the data in the outermost region 
which is located outside of the virial radius ($\timeform{37.8'}-\timeform{42.0'}$).  
We account for the Local Hot Bubble (LHB: $\sim 0.1$ keV with no absorption), the Milky-Way Halo 
(MWH: $\sim 0.3$ keV with absorption) and the CXB component mentioned above. 
The model is described by ${\it  apec+wabs(apec+powerlaw)}$.  
The redshift and abundance in both the {\it apec} components were fixed at 0 and unity, respectively.  
To represent the Galactic and CXB emission, we used the XIS response for a source of uniform brightness.  
The spectrum was well fitted with the above model ($\chi^{2}_{red}$=1.10 for 344 degrees of freedom).  
The temperatures of the LHB and the MWH are $0.11\pm0.01$ keV and $0.37\pm0.03$ keV, respectively, 
consistent with the typical Galactic emission.  
The resultant parameters and the spectrum are shown in Table~\ref{tab:bgd} and in Figure~\ref{fig:bgd}.

\begin{table*}[]
\begin{center}
\caption{
Relative count contributions (\%) due to PSF broadening of the Suzaku XIS mirror. 
Relative contribution (\%) from each sky region in the annular extraction region 
due to the PSF broadening of the Suzaku XIS mirror. Most of the PSF contaminations
in an annular originates from the inner neighboring annular.
}
\begin{tabular}{cccccccccccc}\hline
  Detector/ Sky  & (1) & (2) &(3) &(4)& (5)&(6)&(7)& (8)&(9)\\ \hline
(1)  \timeform{0'}-\timeform{4.2'} & {\bf 90.9} & 8.9 & 0.1 & 0.0 & 0.0 & 0.0 & 0.0 & 0.0 & 0.0 \\
(2) \timeform{4.2'}-\timeform{8.4'}&20.6 & {\bf 75.4} & 3.9 & 0.1 & 0.0 & 0.0 & 0.0 & 0.0 & 0.0 \\
(3)$\timeform{8.4'}-\timeform{12.6'}$ &3.0 & 21.0 & {\bf 71.9} & 3.7 & 0.2 & 0.1 & 0.0 & 0.0 & 0.0 \\
(4)$\timeform{12.6'}-\timeform{16.8'}$ &0.9 & 1.4 & 12.7 & {\bf 75.1} & 9.3 & 0.4 & 0.1 & 0.1 & 0.0 \\
(5)$\timeform{16.8'}-\timeform{21.0'}$ &0.8 & 0.6 & 1.4 & 14.3 & {\bf 73.8} & 8.4 & 0.4 & 0.1 & 0.0 \\
(6)$\timeform{21.0'}-\timeform{25.2'}$ &1.1 & 1.2 & 0.7 & 1.5 & 14.1 & {\bf 72.1} & 8.6 & 0.3 & 0.2 \\
(7)$\timeform{25.2'}-\timeform{29.4'}$ &1.9 & 1.4 & 0.9 & 0.8 & 1.2 & 13.3 & {\bf 72.1} & 7.7 & 0.5 \\
(8)$\timeform{29.4'}-\timeform{33.6'}$ &0.8 & 0.6 & 0.5 & 0.4 & 0.4 & 0.9 & 13.4 & {\bf 73.4} & 9.1 \\
(9)$\timeform{33.6'}-\timeform{37.8'}$ &1.8 & 1.6 & 1.1 & 0.7 & 0.4 & 0.7 & 0.5 & 11.2 & {\bf 71.4} \\ \hline
\label{tab:stray}
\end{tabular}
\end{center}
\end{table*}

\begin{figure*}[htbp]
\begin{tabular}{cc}
\begin{minipage}{0.333\hsize}
(a)\timeform{0'}-\timeform{4.2'}
\\[-0.8cm]
\begin{center}
\includegraphics[angle=-90,scale=0.22]{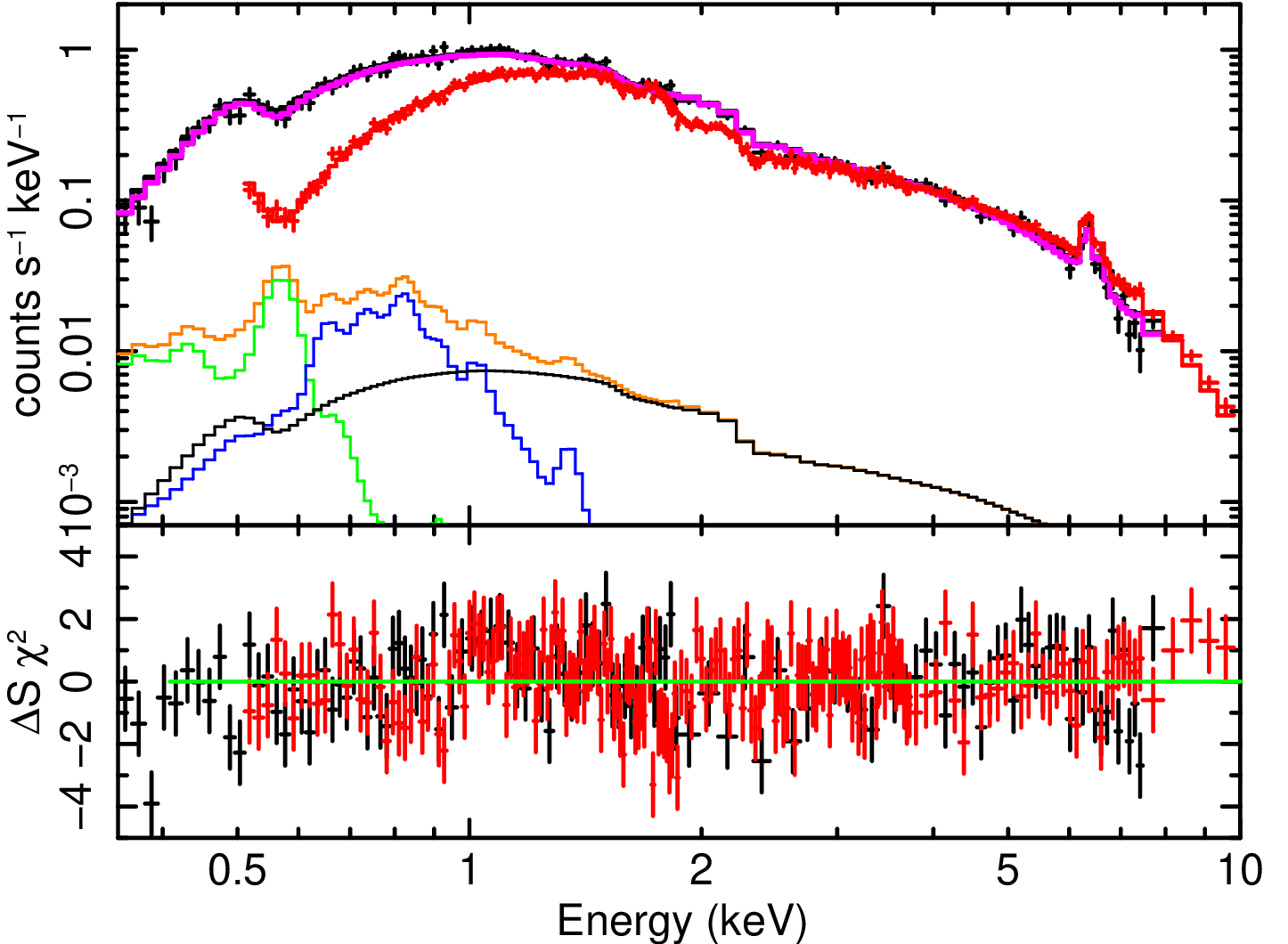}
\end{center}
\end{minipage}
\begin{minipage}{0.3333\hsize}
(b)\timeform{4.2'}-\timeform{8.4'} 
\\[-0.8cm]
\begin{center}
\includegraphics[angle=-90,scale=0.22]{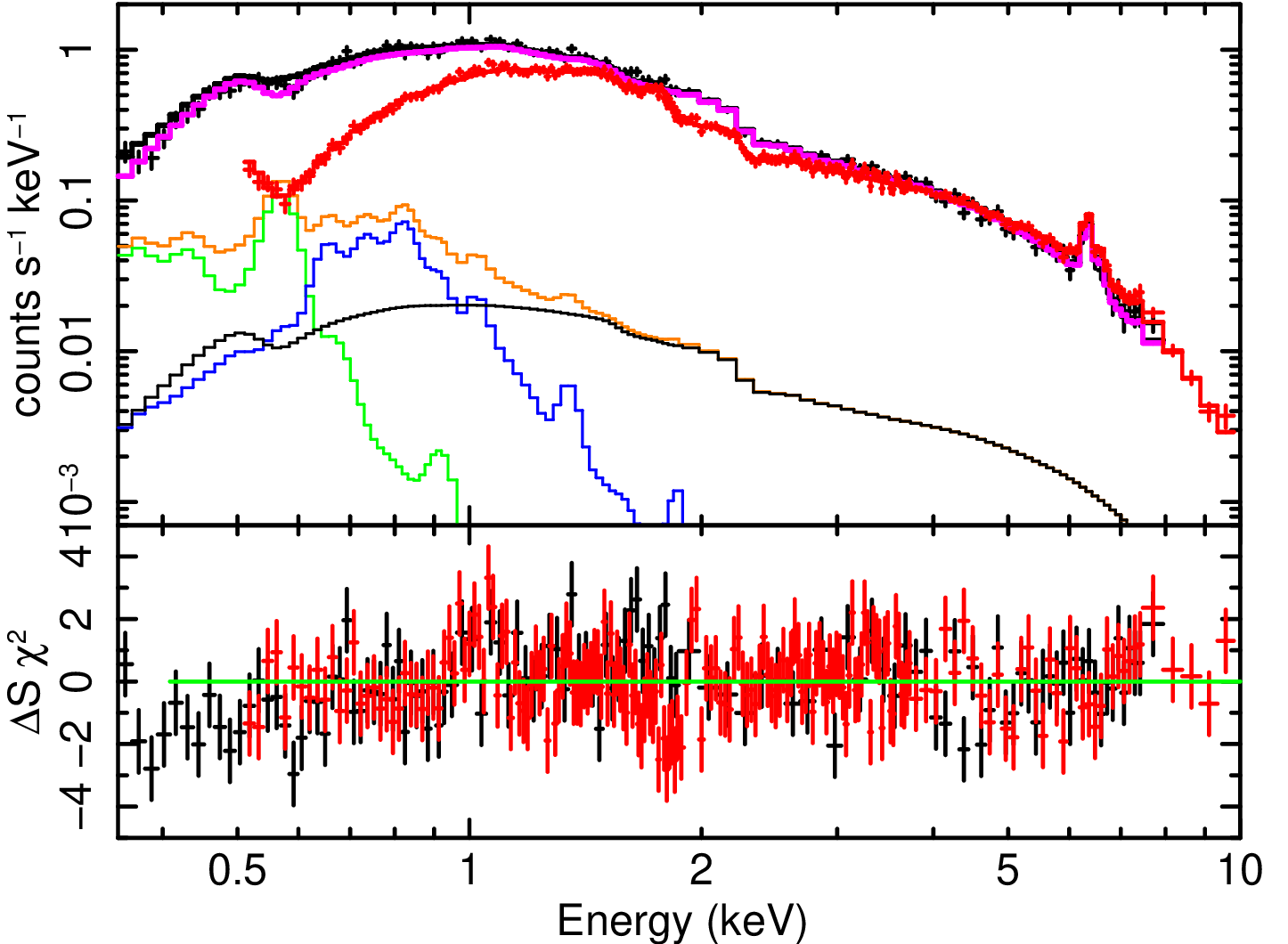}
\end{center}
\end{minipage}
\begin{minipage}{0.3333\hsize}
(c)\timeform{8.4'}-\timeform{12.6'} 
\\[-0.8cm]
\begin{center}
\includegraphics[angle=-90,scale=0.22]{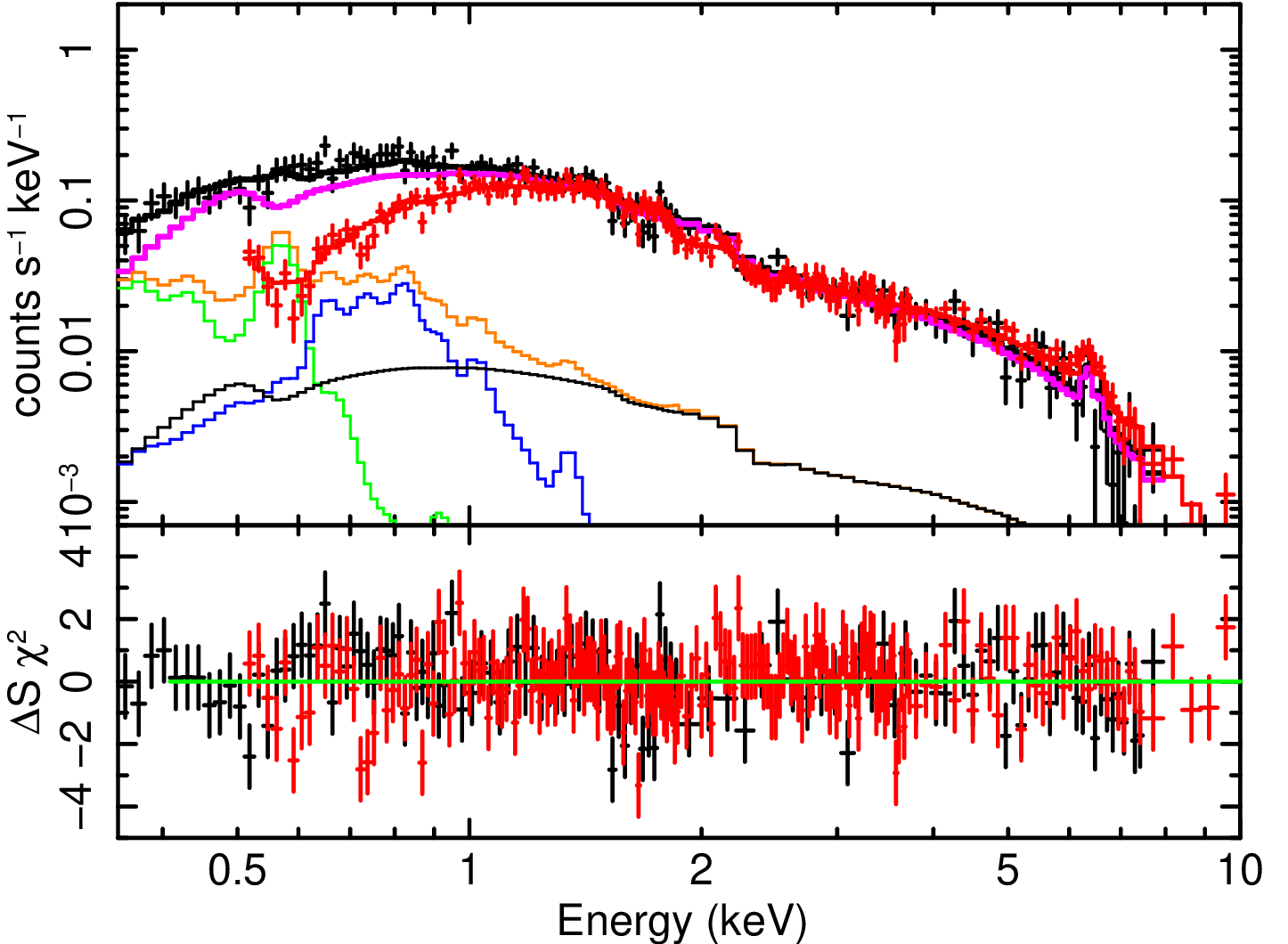}
\end{center}
\end{minipage}\\
\begin{minipage}{0.3333\hsize}
(d)\timeform{12.6'}-\timeform{16.8'} 
\\[-0.8cm]
\begin{center}
\includegraphics[angle=-90,scale=0.22]{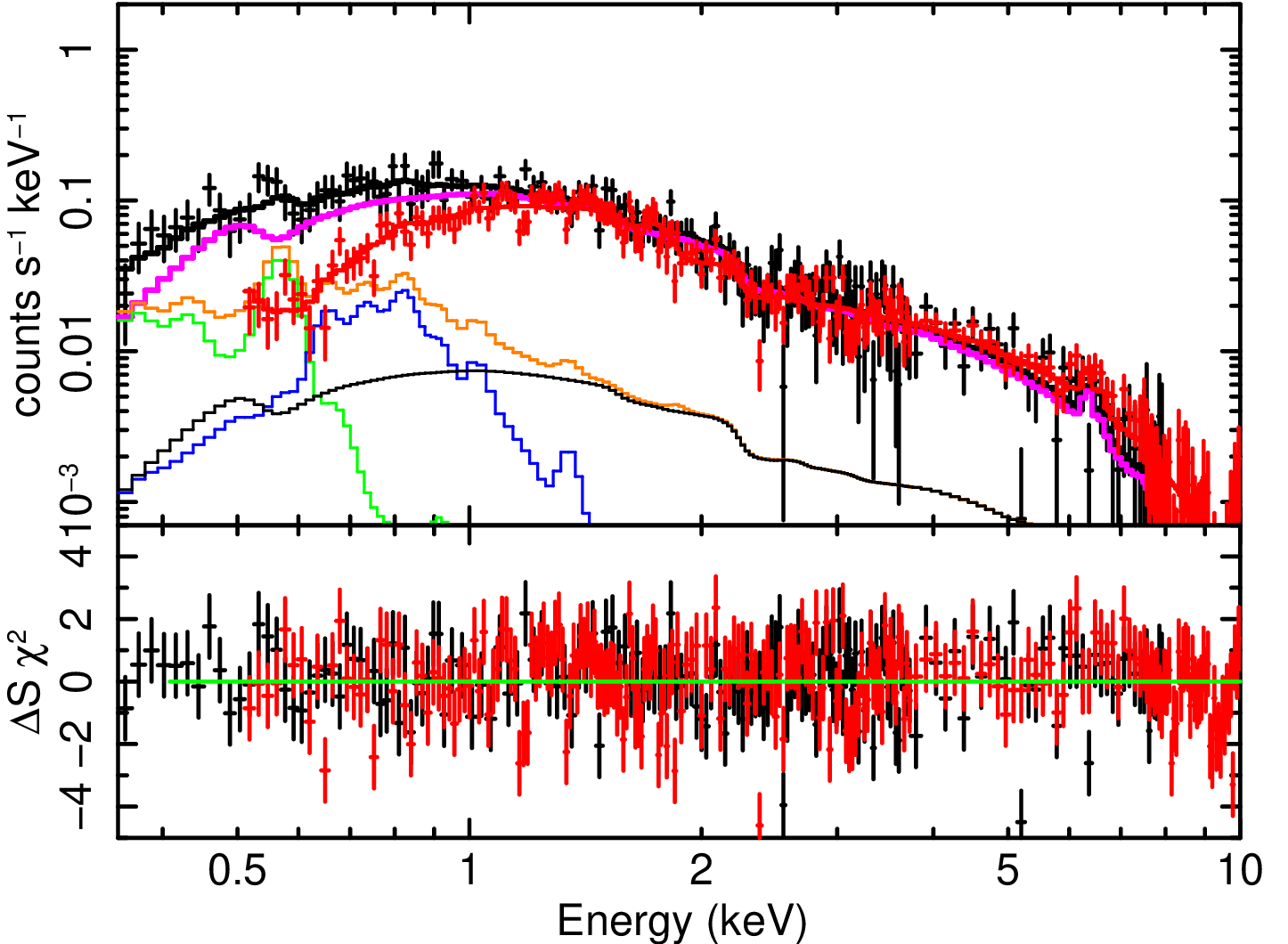}
\end{center}
\end{minipage}
\begin{minipage}{0.3333\hsize}
(e)\timeform{16.8'}-\timeform{21.0'} 
\\[-0.8cm]
\begin{center}
\includegraphics[angle=-90,scale=0.22]{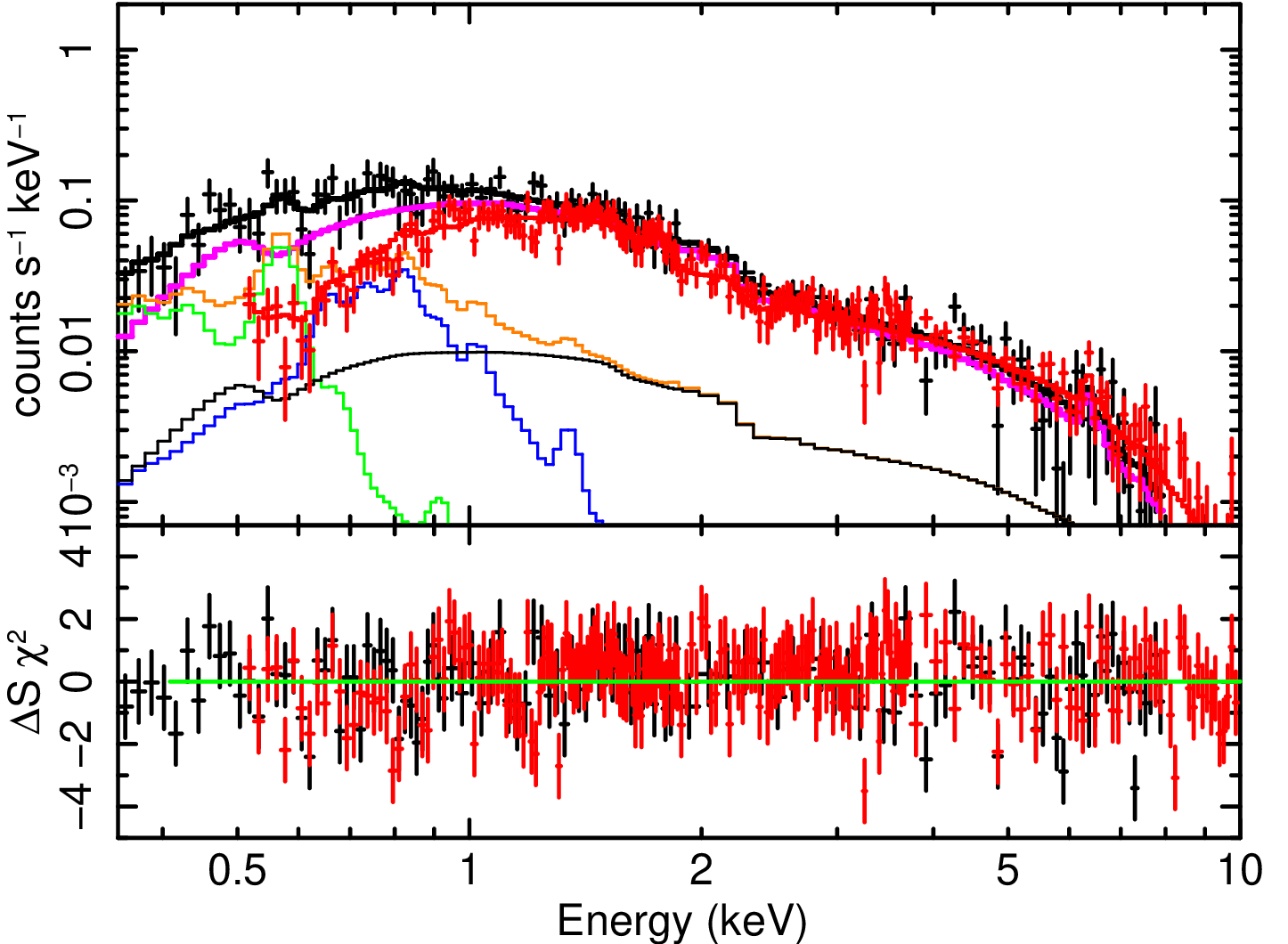}
\end{center}
\end{minipage}
\begin{minipage}{0.33333\hsize}
(f)\timeform{21.0'}-\timeform{25.2'} 
\\[-0.8cm]
\begin{center}
\includegraphics[angle=-90,scale=0.22]{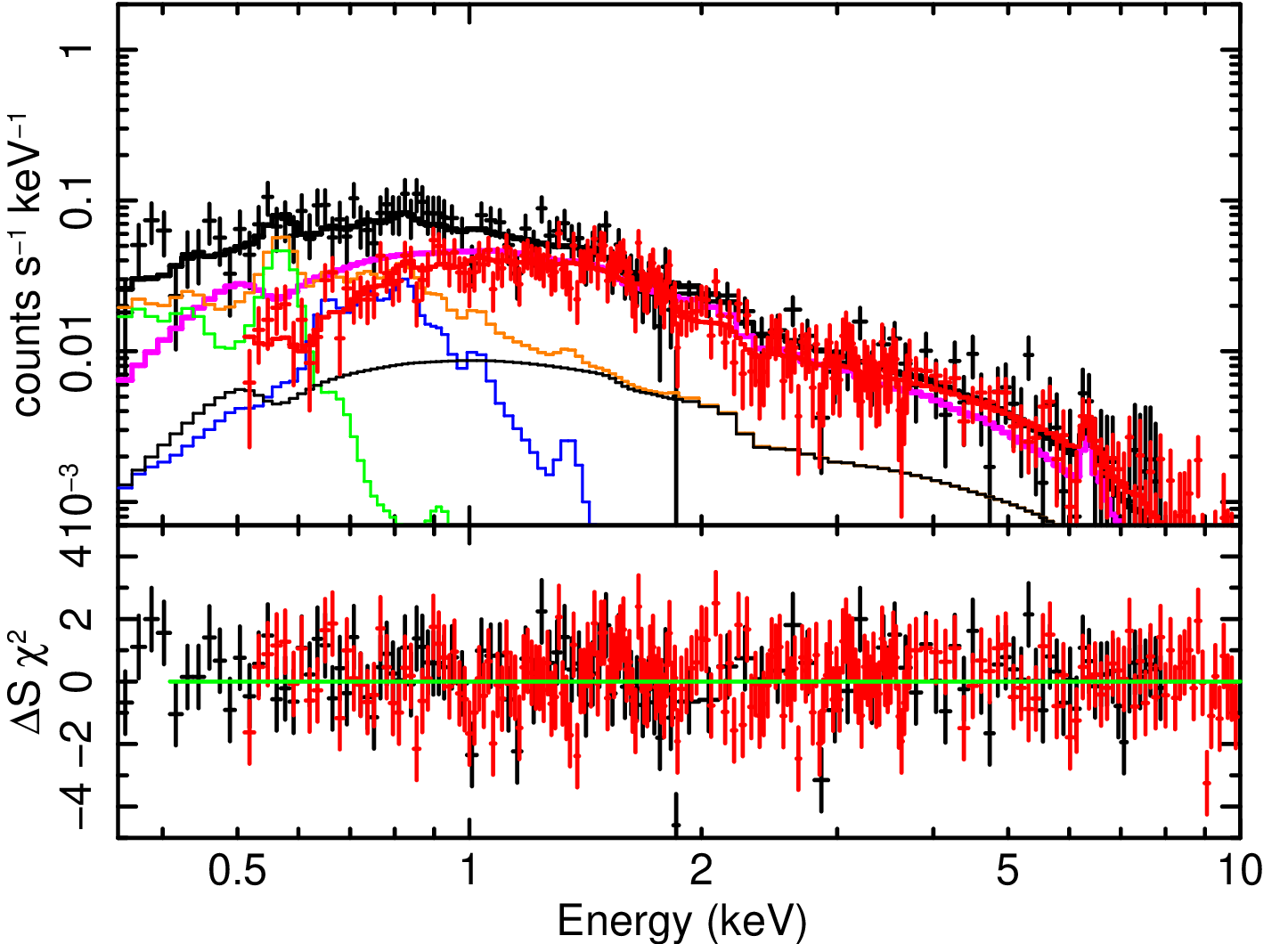}
\end{center}
\end{minipage}\\
\begin{minipage}{0.3333\hsize}
(g)\timeform{25.2'}-\timeform{29.4'} 
\\[-0.8cm]
\begin{center}
\includegraphics[angle=-90,scale=0.22]{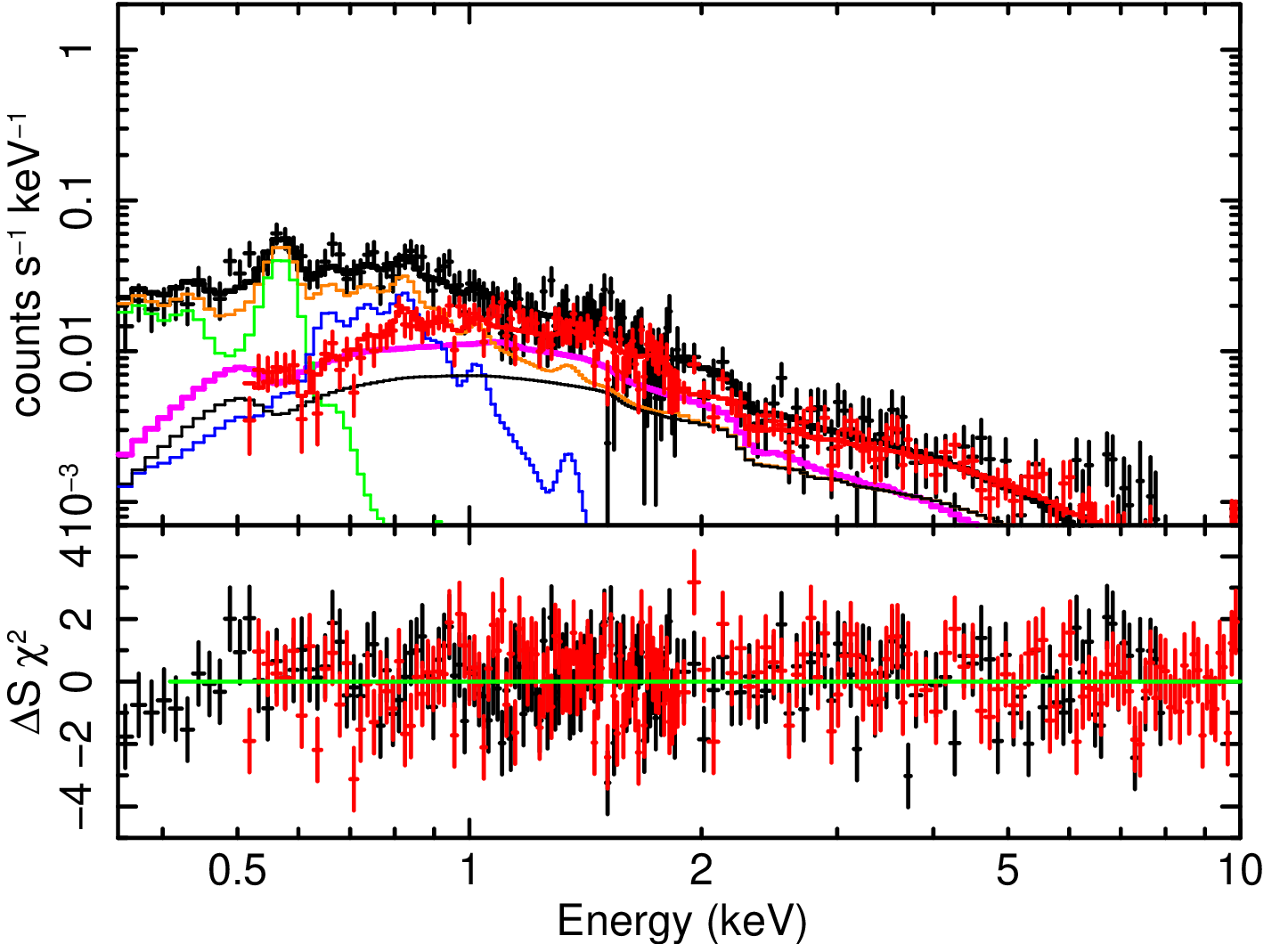}
\end{center}
\end{minipage}
\begin{minipage}{0.3333\hsize}
(h)\timeform{29.4'}-\timeform{33.6'} 
\\[-0.8cm]
\begin{center}
\includegraphics[angle=-90,scale=0.22]{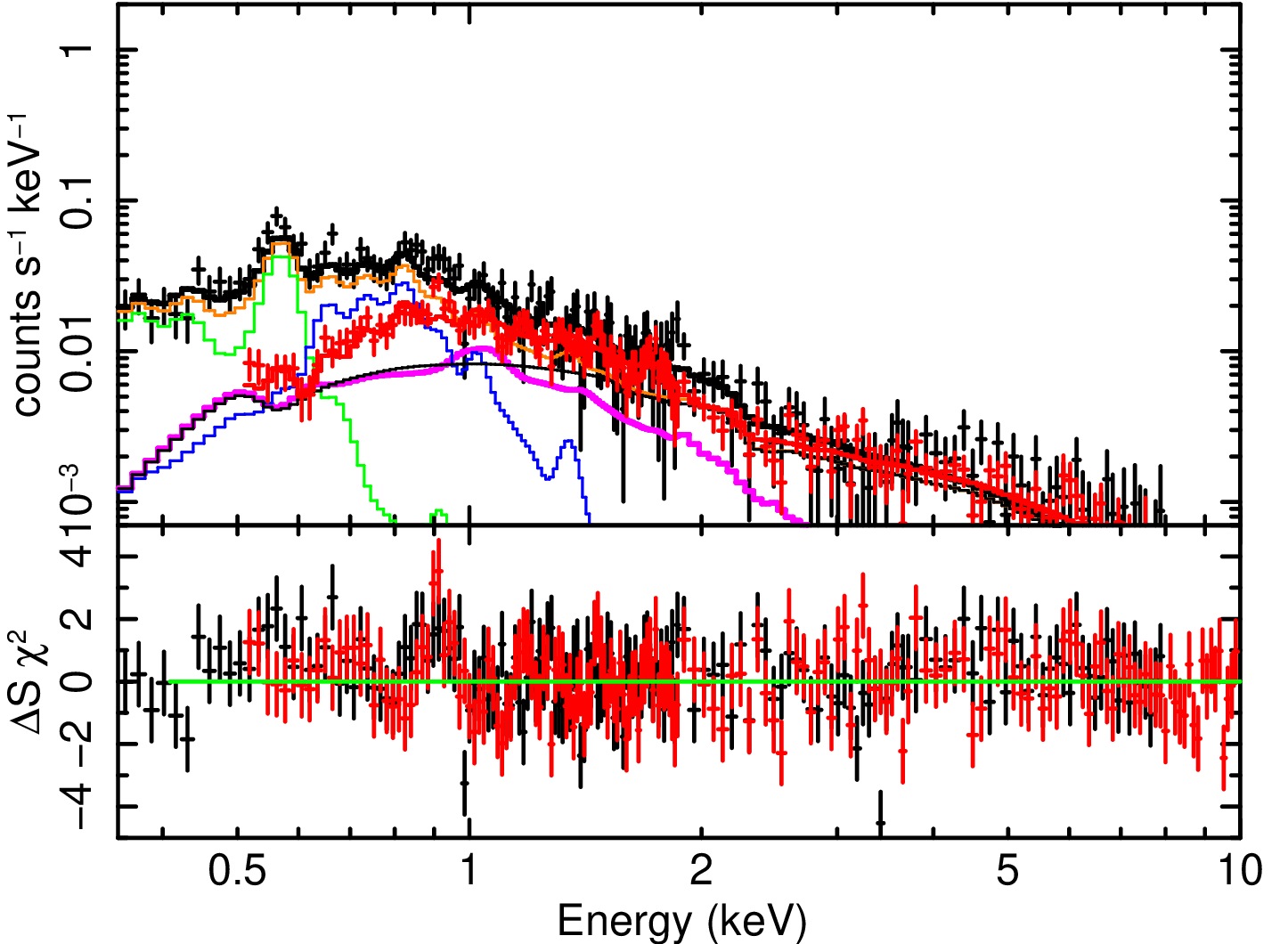}
\end{center}
\end{minipage}
\begin{minipage}{0.33333\hsize}
(i)\timeform{33.6'}-\timeform{37.8'} 
\\[-0.8cm]
\begin{center}
\includegraphics[angle=-90,scale=0.22]{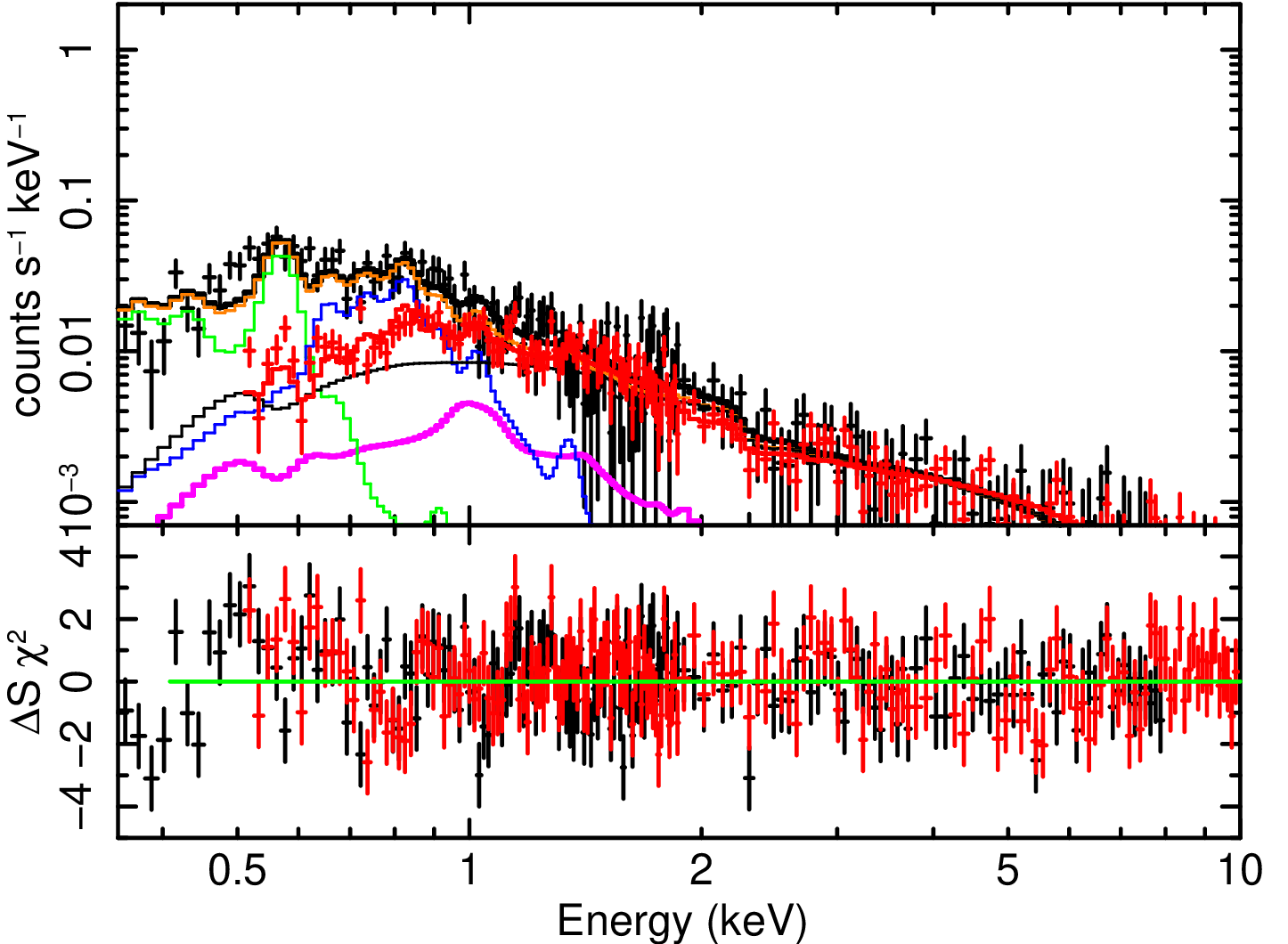}
\end{center}
\end{minipage}
\end{tabular}
\caption{
NXB subtracted spectra in each annular region.  
The XIS BI (Black) and FI (Red) spectra are fitted with the ICM model ({\it wabs + apec}), 
along with the sum of the CXB and the Galactic emission ({\it apec + wabs(apec + powerlaw)}).  
The CXB component is shown with a black curve, and the LHB and MWH emissions are indicated 
by green and blue curves, respectively.  
The total background components are shown by the orange curve. 
}
\label{fig:fit}
\end{figure*}

\subsection{Stray light}
\label{sec:slight}
Stray light consists of photons entering from the detector outside the FOV\@.
The Suzaku optics often show significant effects of stray light from nearby bright X-ray sources \citep{serlemitsos07}.  
The extended point spread function of the telescope with a half-power diameter of
\timeform{1.7'} also causes contamination of photons from nearby sky regions.

\begin{table*}[ht]
\begin{center}
\footnotesize
\caption{Best-fit parameters of the ICM}
\begin{tabular}{cccccccccccc}\hline
  & \timeform{0'}-\timeform{4.2'} &\timeform{4.2'}-\timeform{8.4'}   &$\timeform{8.4'}-\timeform{12.6'}$ &$\timeform{12.6'}-\timeform{16.8'}$ & $\timeform{16.8'}-\timeform{21.0'}$ & $\timeform{21.0'}-\timeform{25.2'}$ & $\timeform{25.2'}-\timeform{29.4'}$ & $\timeform{29.4'}-\timeform{33.6'}$ & $\timeform{33.6'}-\timeform{37.8'}$ \\ \hline
 \multicolumn{10}{c}{NOMIMAL}\\ \hline
$kT$(keV)&
$ {7.14}^{+0.17}_{-0.17} $& 
 $ {6.57}^{+0.13}_{-0.12}  $  & 
 $  {6.93}^{+0.50}_{-0.36}  $  & 
 $  {6.94}^{+0.71}_{-0.50}  $  & 
 $  {6.49}^{+0.65}_{-0.56}  $  & 
 $  {5.70}^{+0.94}_{-0.66}  $  & 
 $  {5.36}^{+1.11}_{-0.91}  $  & 
 $  {1.92}^{+0.39}_{-0.37}  $  & 
 $  {1.58}^{+0.62}_{-0.41}  $  \\
 Z&
  $  {0.29}\pm 0.03 $  & 
 $  {0.33}\pm 0.03  $  & 
 $  {0.26}\pm 0.09  $  & 
 $  {0.20}\pm 0.11  $  & 
 $  {0.22}\pm 0.12  $  & 
 $  {0.26}\pm 0.22  $  & 
 0.2 (FIX)  & 
 0.2 (FIX)  & 
 0.2 (FIX)  \\
norm$^{\ast}$&
 $  {631.2}\pm6.3  $  & 
 $  {209.1}\pm1.8  $  & 
 $  {66.2}\pm1.9  $  & 
 $  {45.7}\pm2.0  $  & 
 $  {27.3}\pm1.4  $  & 
 $  {16.3}\pm1.4  $  & 
 $  {5.0}\pm0.5  $  & 
 $  {2.9}\pm0.4  $  & 
 $  {1.0}\pm0.3  $  \\
 $S^{\parallel}_{\rm 0.4-10 keV}$&
  $  {91.80}\pm0.69  $  & 
 $  {29.54}\pm0.19  $  & 
 $  {10.13}\pm0.66  $  & 
 $  {6.46}\pm0.06  $  & 
 $  {3.67}\pm0.07  $  & 
 $  {2.10}\pm0.05  $  & 
 $  {0.63}\pm0.01  $  & 
 $  {0.14}\pm0.01  $  & 
 $  {0.04}\pm0.01  $ \\
 $\chi^{2}$/d.o.f &
 386 / 311 &
 391 / 311&
 323 / 311  & 
 498 / 404  & 
 371 / 330  & 
 348 / 330  & 
 404 / 347  & 
 388 / 347  & 
 454 / 347  \\ \hline
 \multicolumn{10}{c}{CXB MAX+NXB 3\% ADD}\\ \hline
$kT$(keV)&
  $  {7.11} $  & 
 $  {6.51} $  & 
 $  {6.76} $  & 
 $  {6.91}  $  & 
 $  {6.22}  $  & 
 $  {5.40}  $  & 
 $  {4.10}  $  & 
 $  {0.99} $  & 
 $  {1.05}  $  \\
 Z&
 $  {0.29} $  & 
 $  {0.33} $  & 
 $  {0.25}  $  & 
 $  {0.19}  $  & 
 $  {0.21} $  & 
 $  {0.24} $  & 
 0.2 (FIX)  & 
 0.2 (FIX)  & 
 0.2 (FIX)  \\
 norm&
 $  {612.3}  $  & 
 $  {201.8}  $  & 
 $  {63.6}  $  & 
 $  {46.1} $  & 
 $  {26.0} $  & 
 $  {15.4} $  & 
 $  {4.4}  $  & 
 $  {2.1} $  & 
 $  {0.7}  $  \\
  $S^{\parallel}_{\rm 0.4-10 keV}$&
$  {88.85}  $  & 
 $  {28.47} $  & 
 $  {9.63}$  & 
 $  {6.45}	$  & 
 $  {3.44}	$  & 	
 $  {1.92}	$  & 
 $  {0.47}	$  & 
 $  {0.03}	$  & 
 $  {0.01}	$  \\
 $\chi^{2}$/d.o.f &
  385 / 311  & 
 386 / 311  & 
 326 / 311  & 
 455 / 372  & 
 374 / 330  & 
 349 / 330  & 
 410 / 347  & 
 385 / 347  & 
 479 / 347  \\ \hline
  \multicolumn{10}{c}{CXB MIN+NXB 3\% RED}\\ \hline
 $kT$(keV)&
  $  {7.16}	$  & 
 $  {6.61}	$  & 
 $  {7.07}	$  & 
 $  {7.09}	$  & 
 $  {6.62}	$  & 
 $  {5.95}	$  & 
 $  {6.17}	$  & 
 $  {2.53}	$  & 
 $  {2.25}	$  \\
 Z&
 $  {0.29}	$  & 
 $  {0.33}	$  & 
 $  {0.26}	$  & 
 $  {0.19}	$  & 
 $  {0.22}	$  & 
 $  {0.27}	$  & 
 0.2 (FIX)  & 
 0.2 (FIX)  & 
 0.2 (FIX)  \\
 norm&
 $  {650.2}$  & 
 $  {216.4}$  & 
 $  {69.4}$  & 
 $  {47.0}$  & 
 $  {28.6}$  & 
 $  {17.5}$  & 
 $  {5.8}$  & 
 $  {3.5}$  & 
 $  {1.4}$  \\
 $S^{\parallel}_{\rm 0.4-10 keV}$&
$  {94.93}		$  & 
 $  {30.67}	$  & 
 $  {10.65}	$  & 
 $  {6.66}		$  & 
 $  {3.91}		$  & 
 $  {2.29}		$  & 
 $  {0.78}		$  & 
 $  {0.24}		$  & 
 $  {0.09}		$  \\
 $\chi^{2}$/d.o.f &
  387 / 311  & 
 395 / 311  & 
 323 / 311  & 
 454 / 372  & 
 364 / 330  & 
 349 / 330  & 
 402 / 347  & 
 410 / 347  & 
 450 / 347  \\ \hline
   \multicolumn{10}{c}{CONTAMI 10\% ADD}\\ \hline
 $kT$(keV)&
   $  {6.62}	$  & 
 $  {6.27}		$  & 
 $  {6.75}		$  & 
 $  {6.76}		$  & 
 $  {6.13}		$  & 
 $  {5.46}		$  & 
 $  {5.00}		$  & 
 $  {1.81}		$  & 
 $  {1.31}		$  \\
 Z&
 $  {0.29}		$  & 
 $  {0.32}		$  & 
 $  {0.25}		$  & 
 $  {0.18}		$  & 
 $  {0.22}		$  & 
 $  {0.27}		$  & 
 0.2 (FIX)  & 
 0.2 (FIX)  & 
 0.2 (FIX)  \\ 
norm&
$  {650.2}		$  & 
 $  {214.6}	$  & 
 $  {66.2}		$  & 
 $  {46.3}		$  & 
 $  {27.0}		$  & 
 $  {15.5}		$  & 
 $  {5.2}		$  & 
 $  {3.0}		$  & 
 $  {1.0}		$  \\
  $S^{\parallel}_{\rm 0.4-10 keV}$&
 $  {91.87}	$  &
 $  {29.60}	$  &
 $  {10.10}	$  &	
 $  {6.50}		$  &
 $  {3.61}		$  &
 $  {2.01}		$  &
 $  {0.62}		$  &
 $  {0.14}		$  &
 $  {0.03}		$  \\
 $\chi^{2}$/d.o.f &
 413 / 311  & 
 365 / 311  & 
 322 / 311  & 
 455 / 372  & 
 362 / 330  & 
 347 / 330  & 
 402 / 347  & 
 402 / 347  & 
 450 / 347  \\ \hline
    \multicolumn{10}{c}{CONTAMI 10\% RED}\\ \hline
 $kT$(keV)&
  $  {7.75}		$  &
 $  {6.98}		$  &
 $  {7.17}		$  &	
 $  {7.58}		$  &
 $  {6.94}		$  &
 $  {6.13}		$  &
 $  {5.41}		$  &
 $  {1.93}		$  &
 $  {1.72}		$  \\
 Z&
 $  {0.30}		$  &
 $  {0.33}		$  &
 $  {0.26}		$  &
 $  {0.20}		$  &
 $  {0.23}		$  &
 $  {0.29}		$  &
 0.2 (FIX)  & 
 0.2 (FIX)  & 
 0.2 (FIX)  \\
norm&
$  {606.0}		$  &
 $  {201.8}	$  &
 $  {64.9}		$  &
 $  {44.3}		$  &
 $  {25.4}		$  &
 $  {14.7}		$  &
 $  {5.1}		$  &
 $  {2.9}		$  &
 $  {1.0}		$  \\
 $S^{\parallel}_{\rm 0.4-10 keV}$&
$  {91.04}		$  &
 $  {29.46}	$  &
 $  {10.09}	$  &
 $  {6.49}		$  &
 $  {3.60}		$  &
 $  {2.02}		$  &
 $  {0.63}		$  &
 $  {0.14}		$  &
 $  {0.04}		$  \\
  $\chi^{2}$/d.o.f &
  536 / 311  & 
 521 / 311  & 
 326 / 311  & 
 460 / 372  & 
 392 / 330  & 
 347 / 330  & 
 406 / 347  & 
 388 / 347  & 
 450 / 347  \\  \hline
     \multicolumn{10}{c}{Stray light}\\ \hline
    $kT$(keV)&  
    $  {7.06}		$  &
 $  {7.13}		$  &
 $  {7.26}		$  &
 $  {7.38}		$  &
 $  {6.99}		$  &
 $  {5.83}		$  &
 $  {4.09}		$  &
 $  {0.87}		$  &
 $  {0.87}		$  \\
    Z&
     $  {0.29}	$  &
 $  {0.35}		$  &
 $  {0.28}		$  &
 $  {0.19}		$  &
 $  {0.23}		$  &
 $  {0.28}		$  &
  0.2 (FIX)  & 
 0.2 (FIX)  & 
 0.2 (FIX)  \\
    norm&  
      $  {631.2}	$  &
 $  {210.9}	$  &
 $  {66.8}		$  &
 $  {46.8}		$  &
 $  {27.5}		$  &
 $  {16.7}		$  &
 $  {4.5}		$  &
 $  {2.3}		$  &
 $  {0.9}		$  \\
      $S^{\parallel}_{\rm 0.4-10 keV}$&
       $  {91.13}	$  &
 $  {30.75}	$  &
 $  {10.27}	$  &
 $  {6.58}		$  &
 $  {3.77}		$  &
 $  {2.08}		$  &
 $  {0.44}		$  &
 $  {0.03}		$  &
 $  {0.01}		$  \\
        $\chi^{2}$/d.o.f &
      383 / 311  & 
 526 / 311  & 
 325 / 311  & 
 403 / 330  & 
 367 / 330  & 
 344 / 330  & 
 406 / 347  & 
 399 / 347  & 
 494 / 347  \\
 \hline
 \multicolumn{10}{l}{
*:Normalization of the apec component scaled with a factor
SOURCE-RATIO-REG/$\Omega_e$ from Table~\ref{tab:cxb_fluc},
}\\
\multicolumn{10}{l}{
Norm=$\rm \frac{SOURCE-RATIO-REG}{\Omega_e}\int n_{e}n_{H}
dV/(4\pi(1+z^2)D_{A}^2)\times 10
^{-20}~ cm^{-5}~arcmin^{-2}$,
where $D_A$ is the angular diameter distance to the source.}\\
\multicolumn{10}{l}{
$\parallel$: Photon flux in units of $10^{-6} \rm ~photons~ cm^{-2}~ s^{-1}~ arcmin^{-2}$. Energy band is 
0.4 - 10.0 keV.
}\\
\multicolumn{10}{l}{
Surface brightness of the apec component scaled with a factor
SOURCE-RATIO-REG$\Omega_e$ from Table~\ref{tab:cxb_fluc}.
} 
\label{tab:fit}
\end{tabular}
\end{center}
\end{table*}

To examine the effect of stray light from the bright central region of the A3667 cluster, 
we calculated the photon contributions to our annular regions from different regions of the cluster 
using the ray tracing code {\it xissim}~\citep{ishisaki07}.  
We used version 2008-04-05 of the simulator.  
The surface brightness distribution is one of the input parameters necessary to run {\it xissim}.  
We used a $\beta$-model ($\beta=0.54, r_c =\timeform{2.97'}$) based on the ROSAT
PSPC result as the input X-ray image~\citep{knopp96}.
Table~\ref{tab:stray} shows the resultant photon fractions for each annular region.  
We can see that most ($>70 \%$) of the observed counts come from the ``on-source'' region and 
10--20\% come from the adjacent region which is located on the cluster center side. 
Contribution from the bright central region ($< 12.6'$) is less than 5\% for the annular regions outside of $16.8'$.

\begin{figure*}[t]
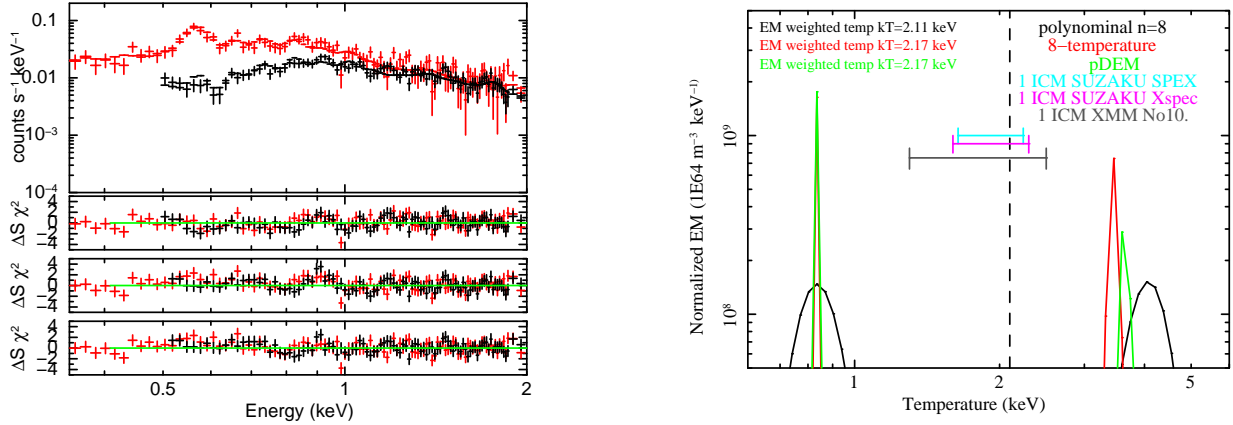

\begin{tabular}{cc}
\begin{minipage}{0.5\hsize}
\begin{center}
\includegraphics[scale=0.3,angle=-90]{DEM_ana.ps}
\end{center}
\end{minipage}
\begin{minipage}{0.5\hsize}
\begin{center}
\includegraphics[scale=0.3,angle=-90]{07t08rv.ps}
\end{center}
\end{minipage}
\end{tabular}
\caption{
(a) Suzaku BI and FI spectrum of the $\timeform{29.4'}-\timeform{33.6'}$ region fitted with the $cie$ model in SPEX.
 To make it easy to see the "1 keV bump", we show the 0.35-2.0 keV spectrum.
 The lower panels show the residuals of a $cie, 2cie, pdem$ model fit to spectrum.
(b) Emission Measure  distribution. Each color shows the results of a different DEM model.
All models show a twin EM peak around 0.9 keV and 4 eV, and the EM weighted temperature is 2.1 keV, 
agrees with 
1 k$T$ model of the Suzaku (Mgenta: 1.92 keV,  
$\timeform{29.4'}-\timeform{33.6'}$) and XMM (Gray: 1.9 keV for region 10, \cite{finoguenov10}) results.
}
\label{fig:dem}
\end{figure*}

\begin{table*}[t]
\begin{center}
\caption{Fit results for  the $\timeform{29.4'}-\timeform{33.6'}$ region with 5 different thermal models. } \label{tab:dem}
\begin{tabular}{cccccccc}\hline 
 Method 	& temperature				(keV)				&norm$^\ast$	&	$\chi^{2}$ /d.o.f\\ \hline 
$1 kT$		&	$1.92\pm0.38$ &		1.23$\pm0.11$				&388/347 \\
$2 kT$		&	0.89$\pm0.09$, 	 4.42$\pm2.7$&	0.69$\pm0.27$, 0.57$\pm0.33$	&360 /345\\ 
$pdem$ 	& --	& 	--	&368/334\\ 
$dem^{\ast}$(poly:n=8)	&--	&	--	 &96/89\\
$dem^{\ast}$(8T)&--	&	--	&114/81\\
\hline
 \multicolumn{4}{l}{$\ast$: The $dem$ can be used only as an independent model,
 }\\
  \multicolumn{4}{l}{
 therefore we subtracted all background components.}
\end{tabular}
\end{center}
\end{table*}

\section{Results }\label{sec:specana}
\subsection{Spectral fitting}
For spectral fitting, we estimate the effective area for each annular using $xissimarfgen$ \citep{ishisaki07}
 assuming a $\beta$-model ($\beta$=0.54, $r_{c}$=\timeform{2.97'}:\cite{knopp96}) 
 surface brightness profiles for the ICM of A3667.
We carried out spectral fitting for the individual annular regions separately.

We modeled the spectrum in each annulus as the sum of the ICM and the
sky background components. Intensities, photon indices and temperatures
of the CXB and the Galactic emission were all fixed at the values from table \ref{tab:bgd}.  
We employed a single temperature thermal model ($wabs\times apec$) for the ICM emission of A3667\@.  
The estimated NXB spectrum was subtracted from each data set before the spectral fit.  
The interstellar absorption was kept fixed using the 21cm measurement
of the hydrogen column, $N_{\rm H}=4.7 \times 10^{20}\rm cm^{-2}$ \citep{dickey90}.  
The solar abundance in our analysis was defined by \citet{anders89}.  
We used XSPEC ver12.4.0 for the spectral fit.

In the central regions within $\timeform{25.2'}$, the free parameters were
temperature $kT$, normalization {\it norm} and metal abundance $Z$ of the ICM component.  
In the outer regions ($>\timeform{25.2'}$), we fixed the ICM metal abundance to 0.2, 
which is the typical value observed in cluster outskirts \citep{fujita08}.  
We used the energy ranges 0.35-8 keV for BI and 0.5-10 keV for FI, respectively.  
The relative normalization between the two sensors was a free parameter in this fit
to compensate for cross-calibration errors.  
The spectra and the best-fit models for all the annular regions are shown in Fig~\ref{fig:fit}.  
The parameters and the resultant $\chi^2$ values are listed in Table~\ref{tab:fit}.

We also examined the effect of systematic errors on our spectral parameters.  
We considered the systematic error for the NXB intensity to be $\pm 3 \%$~\citep{tawa08}, 
the fluctuation of the CXB is shown as CXBMAX and CXBMIN, and 
the systematic error in the OBF contamination is considered to be $\pm 10 \%$
CONTAMI+10\% and CONTAMI-10\% \citep{koyama07} in Table~\ref{tab:cxb_fluc}, respectively.

To examine the influence of stray light on our spectral results,
we produced mock stray light spectra using {\it xissim} and fitted the annular spectra after this stray light spectrum was subtracted.
To simulate the stray light spectrum by {\it xissim}, the spectral shape and the surface brightness distribution of the ICM
are the input parameters.
We assumed the uniform spectrum that was observed in the central region (\timeform{0'}-\timeform{4.2'}: 
$kT\sim 7.0$ keV) and the $\beta$-model surface brightness distribution described above.
The resultant stray light corrected best fit values are shown in in table \ref{tab:fit} in ``stray light'' row.
We confirmed that the parameters after correction for the stray light effect stayed
within the statistical and systematic errors of our spectral fit.

\begin{figure*}[]
\begin{tabular}{cc}
\begin{minipage}{0.5\hsize}
(a)Temperature
\\[-0.8cm]
\begin{center}
\includegraphics[scale=0.33,angle=-90]{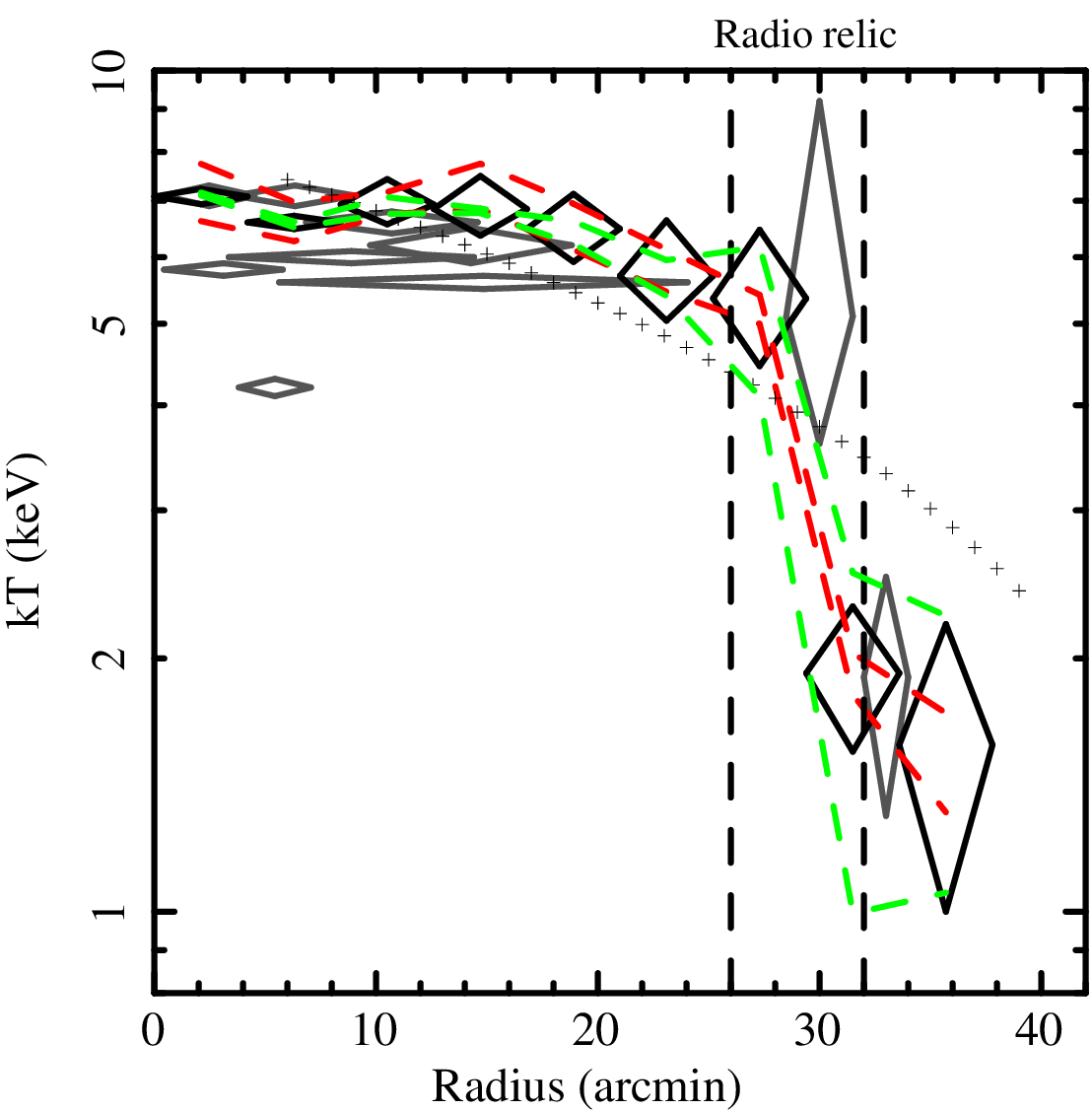}
\end{center}
\end{minipage}
\begin{minipage}{0.5\hsize}
(b) Surface brightness
\\[-0.8cm]
\begin{center}
\includegraphics[scale=0.33,angle=-90]{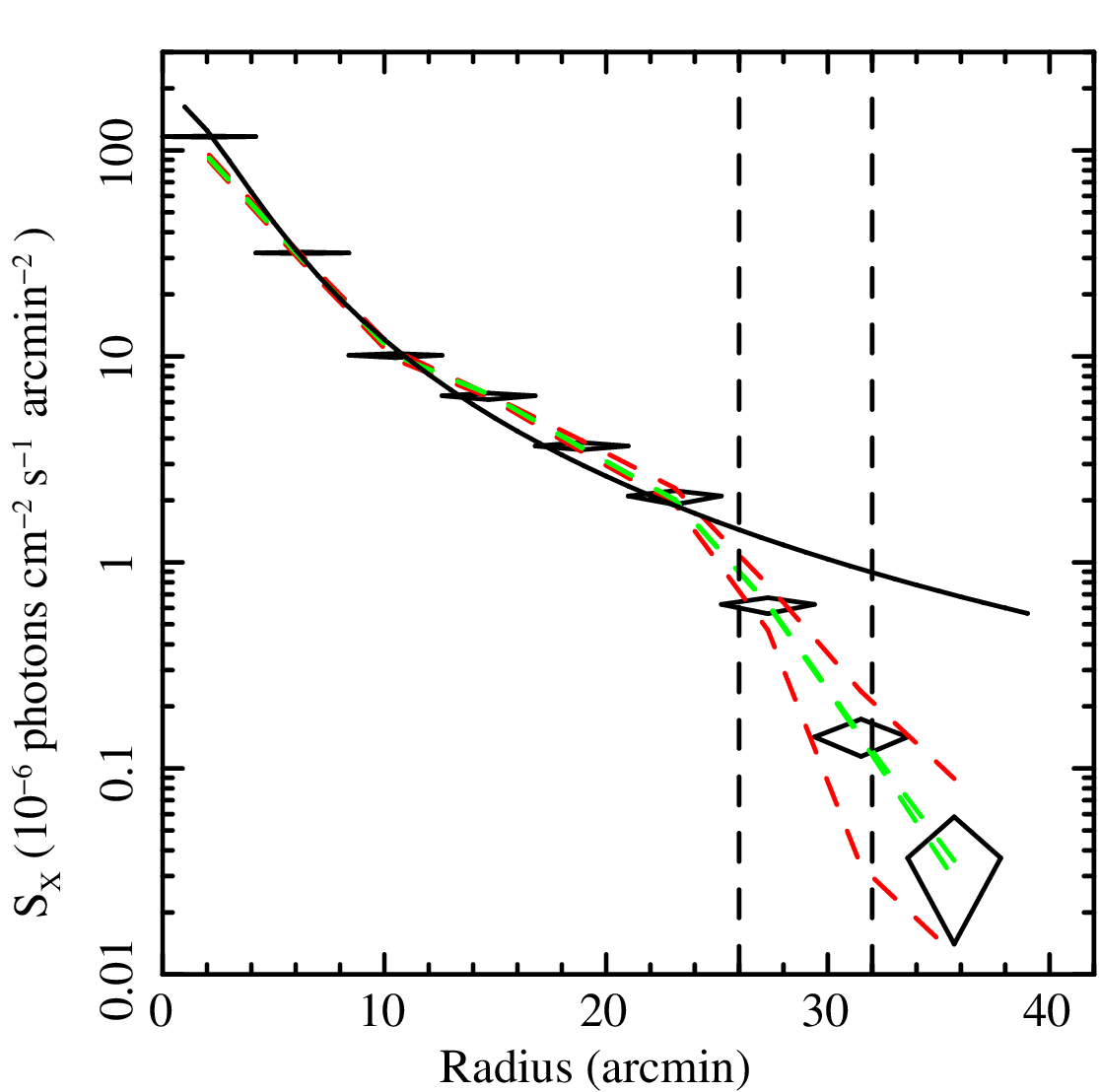}
\end{center}
\end{minipage}\\
\begin{minipage}{0.5\hsize}
(a)Abundance
\begin{center}
\includegraphics[scale=0.33,angle=-90]{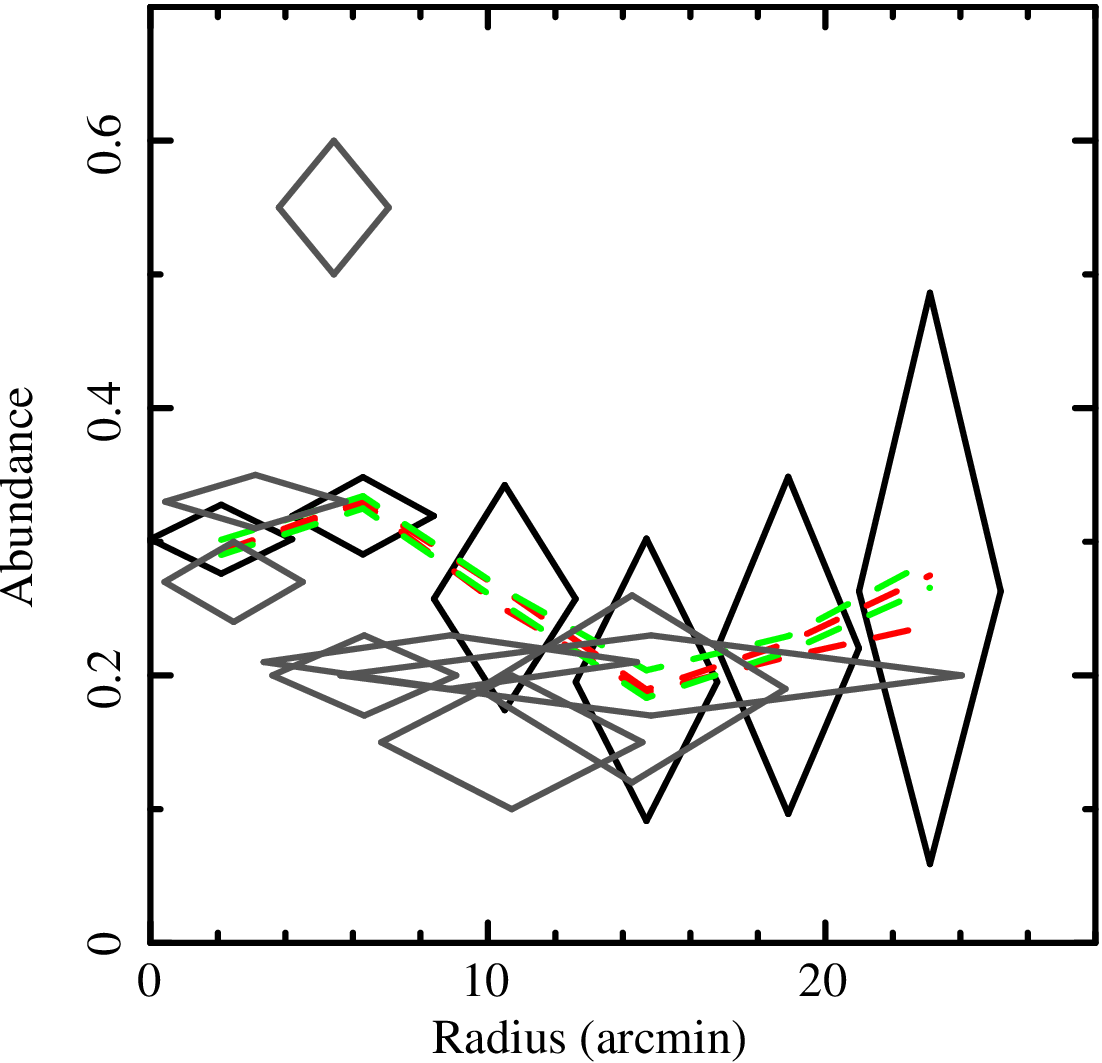}
\end{center}
\end{minipage}
\begin{minipage}{0.5\hsize}
(d)Deprojected electron density 
\begin{center}
\includegraphics[scale=0.33,angle=-90]{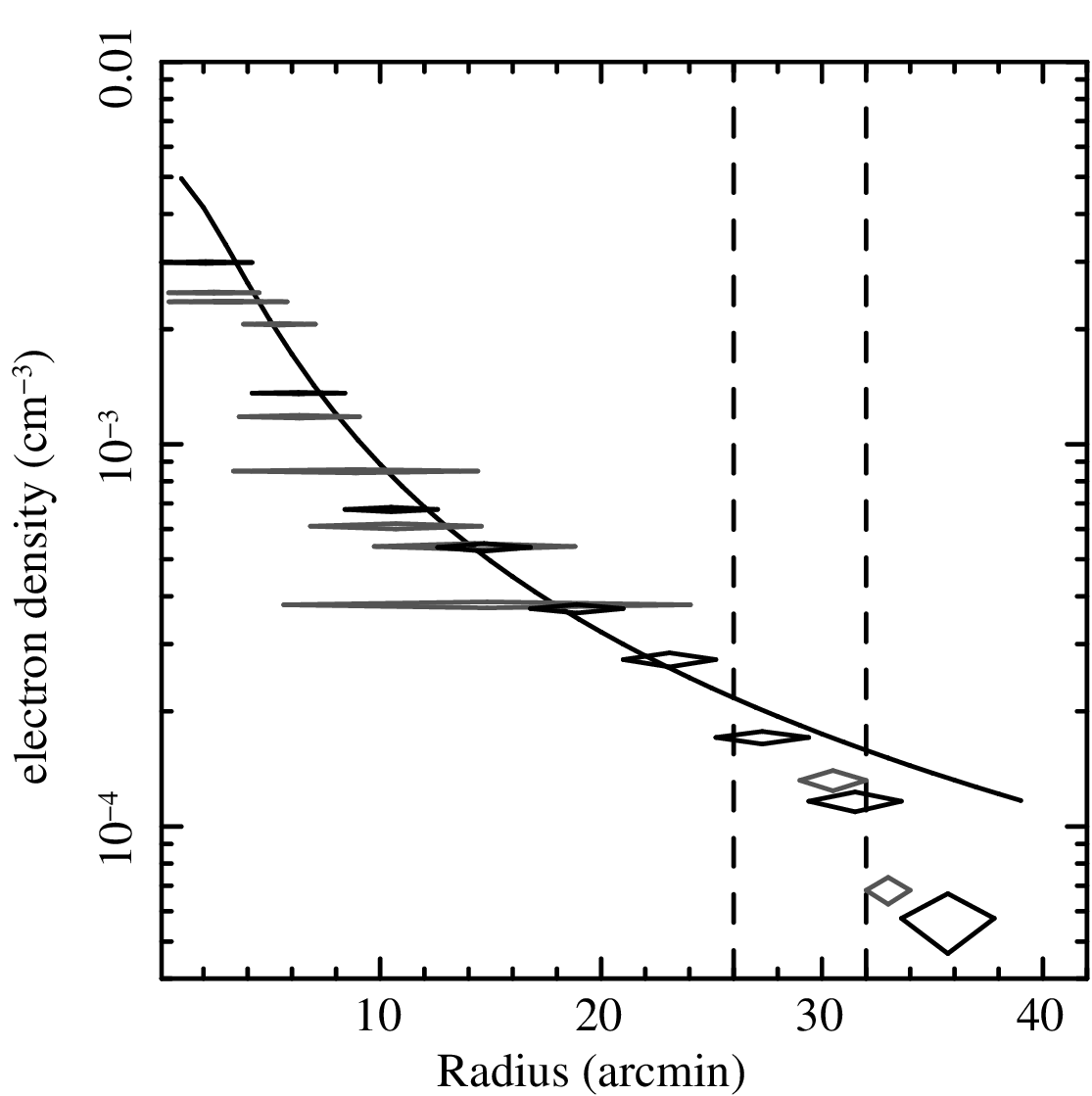}
\end{center}
\end{minipage}
\end{tabular}
\caption{\small
Radial profiles of (a) ICM temperature, (b) Surface brightness,
(c)Abundance, (d)Deprojected electron density.
The Suzaku best fit values are shown by black diamonds. 
In temperature profiles, gray diamonds show the results of XMM-Newton \citep{briel04,finoguenov10}.
Black dashed vertical lines show the approximate radial boundaries of the northwest radio relic. 
The Green and red dash lines show typical systematic changes of the best fit 
values due to changes of OBF contaminants and the NXB level.
In temperature profiles, the crosses show an average profile given by \citet{pratt07} for 
A3667.
In the surface brightness and deprojected electron density profiles, 
the black solid curve shows 
the $\beta$-model obtained by ROSAT PSPC \citep{knopp96}.
}
\label{fig:profile}
\end{figure*}

\subsection{Spectral features near the radio relic}\label{sec:dem}
In the edge region of the radio relic
($\timeform{29.4'}-\timeform{33.4'}$), the single temperature model
for the ICM does not give a good fit in the low energy range. 
As shown in Fig.~\ref{fig:fit}(h), there is a residual feature around 0.9 keV
which was already reported from Suzaku and XMM-Newton observations~(\cite{nakazawa09,finoguenov10}).  
In this section, we look into the possibility of a multi-temperature plasma.

To test the multi-temperature structure, we first included an
additional thermal component ($2 kT$ model) to the model. 
The metal abundances were fixed at 0.2 times solar for both thermal models. 
We then obtained an almost acceptable fit with $\chi^2/d.o.f.= 360/345$.  
The resultant high and low temperatures are $kT_{\rm high}=4.42\pm2.7$ keV
and $kT_{\rm low}=0.89\pm0.09$ keV, respectively. 
The single temperature fit gave $kT = 1.92$ keV, which was in the middle of these two temperatures.

Additionally, we adopted differential emission measure (DEM) models using the SPEX spectral fitting package~\citep{kaastra96}.  
Several DEM models can account for complex emission measure (EM) distributions,
and here we tried 3 models.  
The first DEM model we employed is a polynomial differential emission measure $pDEM$ model.  
The second model is a polynomial method 
as $dem$ (poly), 
which has been applied to several sources \citep{lemen89,stern95}.  
In this method, the logarithmic emission measure is described as the sum of n Chebyshev polynomials as a
function of logarithmic temperature.  
The last one is a Multi-temperature method
as $dem$ (multi T)
, which is a sum of  gaussian temperature distributions
;
\begin{equation}
EM_{\rm total}=\sum_i\frac{EM_{i}}{\sqrt{2\pi}\sigma_{i}}\rm exp[-( log({\it T})-log({\it T_{i}}))^{2}/2\sigma_{i}^{2}]
\end{equation}
where $EM_{i}$ is the total emission measure of the component $i$, and
$T_{i}$ is the central temperature.

Fig~\ref{fig:dem}(a) shows the result of spectral fits with the $pDEM$ model.  
The lower panel shows the residuals by fitting with $1 kT, 2kT$, and $pDEM$ models.  
In the $ 2kT $ and $pDEM$ cases, the residuals around 0.9 keV are less significant than for the $1kT$ case.
Fig~\ref{fig:dem}(b) shows the resultant EM distribution based on the DEM analysis.  
In this analysis, we fixed the metal abundance to 0.2 solar (with the solar abundance table:~\cite{anders89}) and 
the energy range from 0.5 to 5.0 keV\@.  
The result shown in Table~\ref{tab:dem}.
The resultant emission measure distribution indicates 2 peaks around 0.9 and 4 keV, which is very similar to the result of a $2kT$ fit.

The emission measure weighted temperature becomes 2.1 keV, which is
consistent with the $1 kT$ results obtained with Suzaku and XMM-Newton~\citep{finoguenov10}.  
In ~\citet{finoguenov10}, most of the outer regions (No.\ 1, 14 in Table~1) showed $kT\sim 1$ keV
consistent with the results of the present analysis.
\citet{nakazawa09} also reported a 0.9 keV component at the radio
relic region using Suzaku data.  
Those results suggest either that (1) we observed the foreground Galactic component or that 
(2) the $\timeform{29.4'}-\timeform{33.6'}$ region is a mixture of low and high temperature ICM\@.

As discussted by \citet{kaastra04}, a multi-phase gas makes an average
temperature spectrum which is difficult to be distinguished by current
X-ray satellites.  To solve this type of problem by line diagnosis, a
future X-ray satellite with a large grasp (FOV$\times$ Effective area)
and good energy resolution is highly 
desirable.
\subsection{ICM properteis}
\label{sec:temp}
Recent studies from Suzaku have revealed the temperature structure in the outskirts of relaxed clusters, 
and suggested that ion-electron temperature equilibrium was not reached there \citep{ george08,reiprich09,hoshino10}.
In the merging cluster Abell 2142, the temperature profile suggests some excess compared with relaxed clusters~\citep{akamatsu11}.  
We have relatively little information about the ICM properties in the outer regions of merging
clusters, However, our results on A3667 give us new information. 
Here, we use the ICM parameters based on our spectral fits shown in table~\ref{tab:fit}.

Figure~\ref{fig:profile}(a) shows the radial profile of the ICM temperature in A3667\@.  
The temperature within $\timeform{15'}$ (960 kpc) from the cluster center shows a fairly constant value of $\sim 7$ keV, 
and it gradually decreases toward the outer region down to $\sim 5$ keV around the radio relic ($\sim \timeform{27'}$).  
As mentioned in Section~\ref{sec:intro}, ~\citet{finoguenov10} reported a significant
drop of temperature and surface brightness in the radio relic region.
Our temperature profile also shows a remarkable drop by a factor of
2--3 in this region even considering the systematic errors.

We compare our temperature profile with the previous XMM-Newton result for A3667~\citep{briel04,finoguenov10} 
and an average profile given
by \citet{pratt07}.  The average profile was made from scaled
temperatures for a sample of 15 clusters, and it is described by the
following formula
\begin{eqnarray}
T/T_X = 1.19 - 0.74R/R_{200},
\label{eq:pratt}
\end{eqnarray}
where $T_X$ is the average temperature and $R_{200}$ is the virial
radius.  We used $T_X=7.0$ keV and $R_{200}=\timeform{34.1'}$ for A3667\@.  
The temperature profile of A3667 shows good agreement with
the previous XMM result for the radius range within the radio relic, as shown in Fig.~\ref{fig:profile}(a).  
The observed temperatures of A3667 are somewhat higher than the average levels between $10'$ and $26'$. 
Around the radio relic region, the average profile predicts 4.0 keV but the observed temperature is 5.3 keV, 
which is in excess by about 30\%.

The surface brightness profile shows a good agreement with the
$\beta$-model ($\beta=0.54$) up to the radio relic region.  As shown
in Fig.~\ref{fig:profile}(b), the surface brightness also shows a drop
by an order of magnitude across the relic.
Figure~\ref{fig:profile}(c) shows the metal abundance profile of A3667, which shows a drop from 0.3 to 0.2 solar. 
Such a drop is seen in other merging or non-merging clusters based on previous studies by XMM-Newton ~\citep{matsushita11,lovisari09}.  
The abundance profile 
shows
 a significant excess around $0.1 r_{200}$ 
possibly due to the effect of the cold front.

We calculated the deprojected electron density profile, based on the surface brightness results.  
The resultant profile is shown in Fig.~\ref{fig:profile}(d) along with the previous XMM results~\citep{briel04, finoguenov10}, 
and the $\beta$-model profile from ROSAT PSPC ($\beta=0.54, r_c =\timeform{2.97'}$).  
Our electron density profile and the XMM-Newton results are both in good agreement with the $\beta$-model 
profile up to the radius of the radio relic.

Although the surface brightness of A3667 shows no significant excess over the $\beta$ model, 
the temperature profile is flatter and higher than the average profile.  
A similar feature was already reported for 
another
 merging cluster, A2142~\citep{akamatsu11}.  
In A2142, the temperature profile along the merger axis shows slightly faster decline compared with 
the average temperature curve for other relaxed systems studied with Suzaku. 
In the A3667 case, it is opposite because the temperature shows a slower decline than in other relaxed clusters.  
The ROSAT and XMM images show an elongated shape along the merger axis direction.  
The temperature excess and elongated morphology suggest that the ICM is heated by the merger
event and the time elapsed since then is not enough for full relaxation.  
However, the ICM properties are only measured in the direction of the merger axis with this accuracy. 
Comparison with the ICM properties in the 
perpendicular
 direction would be important.

\section{Discussion}
\label{sec:discussion}
Suzaku performed 3 pointing observations of Abell 3667 along its merger axis.  
The temperature, surface brightness, abundance, and deprojected electron density profiles 
were obtained up to 0.9 times the virial radius ($r_{200}=2.26$ Mpc).  
The temperature and electron density show a significant ``drop'' around the radio relic region,
whose feature suggest a shock front.  
We evaluate the cluster properties (shock wave and entropy) and discuss their implications below.

\subsection{Possibility of a shock at the radio relic region}
\label{sec:shock}
Recent X-ray studies with Chandra showed clear evidence of a shock in 1E~0657-56 and A520~\citep{clowe06,markevitch05}.  
However, there is still little information about the nature of shocks in clusters.
Recently, \citet{finoguenov10} reported a sharp edge in the surface brightness at the outer edge of the NW radio relic in A3667\@.  
They interpreted the feature in terms of a shock, and derived the Mach number ${\cal M}\sim 2.44\pm0.77$ and $\sim 1.68\pm0.16$, based on the jumps of temperature and electron density in the radio relic region, respectively.

\begin{figure*}[t]
\begin{tabular}{cc}
\begin{minipage}{0.5\hsize}
(a)Puressure
\begin{center}
\includegraphics[scale=0.35,angle=-90]{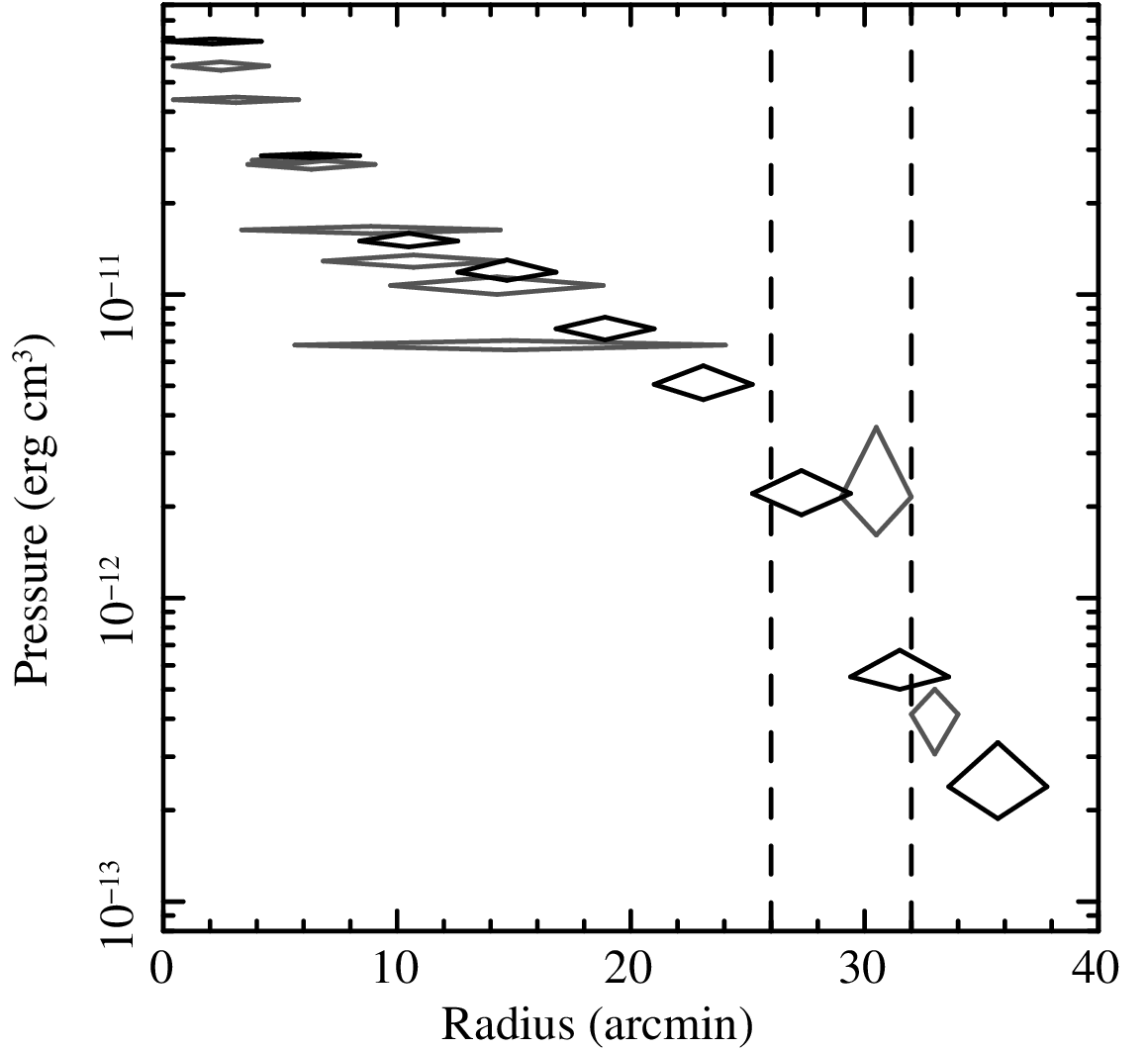}
\end{center}
\end{minipage}
\begin{minipage}{0.5\hsize}
(b)Entropy
\begin{center}
\includegraphics[scale=0.35,angle=-90]{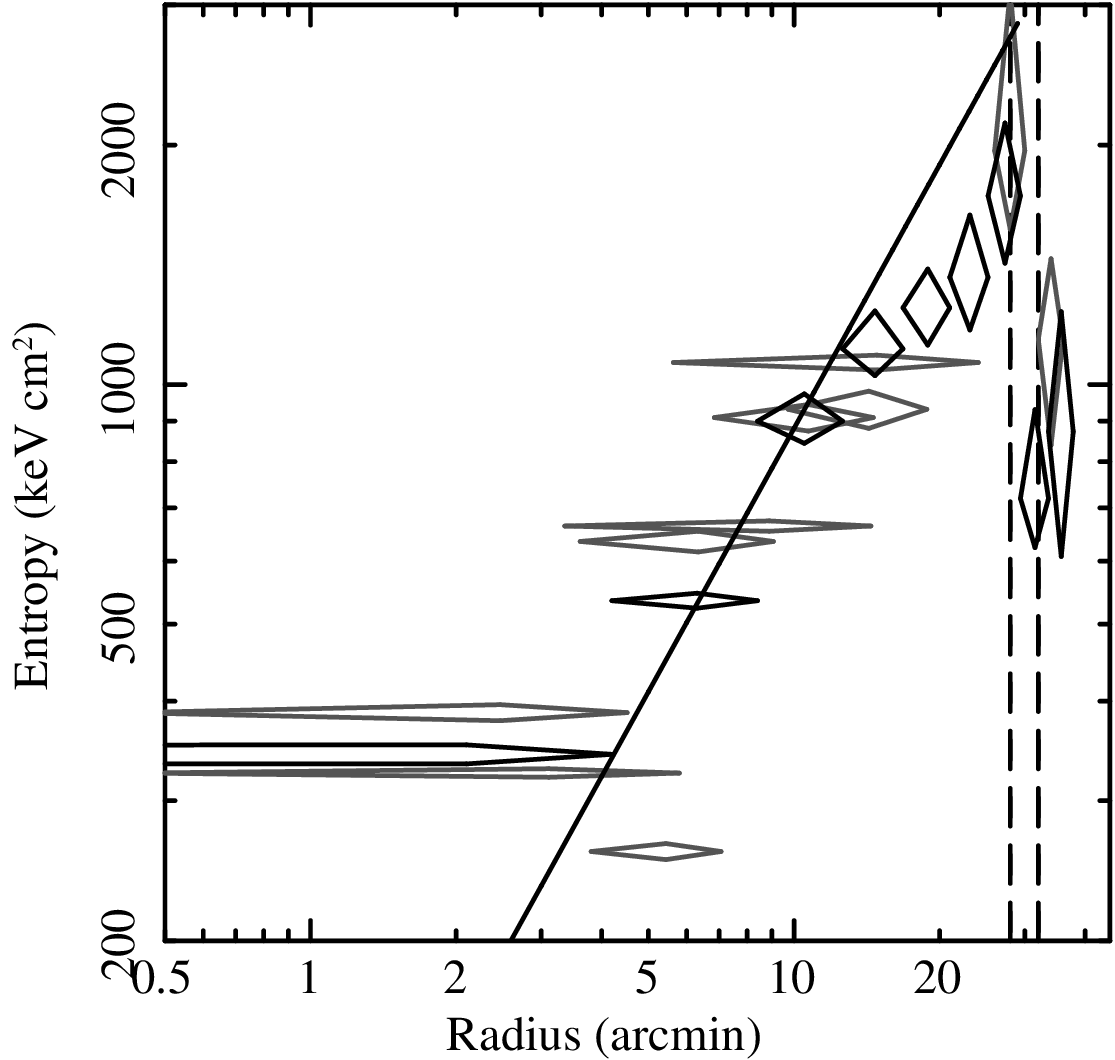}
\end{center}
\end{minipage}\\
\begin{minipage}{0.5\hsize}
(c)Entropy ratio to $r^{1.1}$
\begin{center}
\includegraphics[scale=0.35,angle=-90]{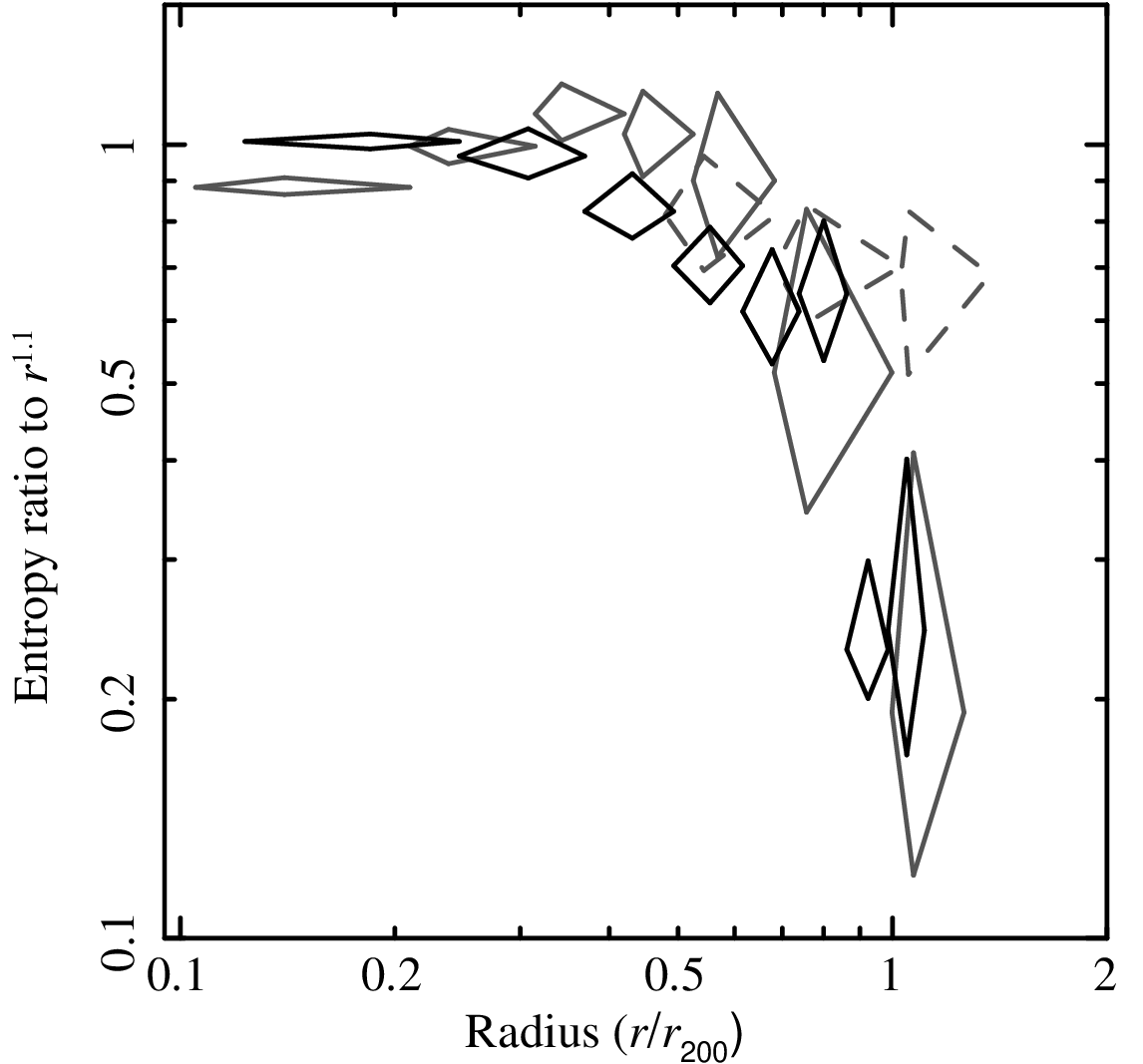}
\end{center}
\end{minipage}
\begin{minipage}{0.5\hsize}
(d)Equilibrium time after shock heating
\begin{center}
\includegraphics[scale=0.35,angle=-90]{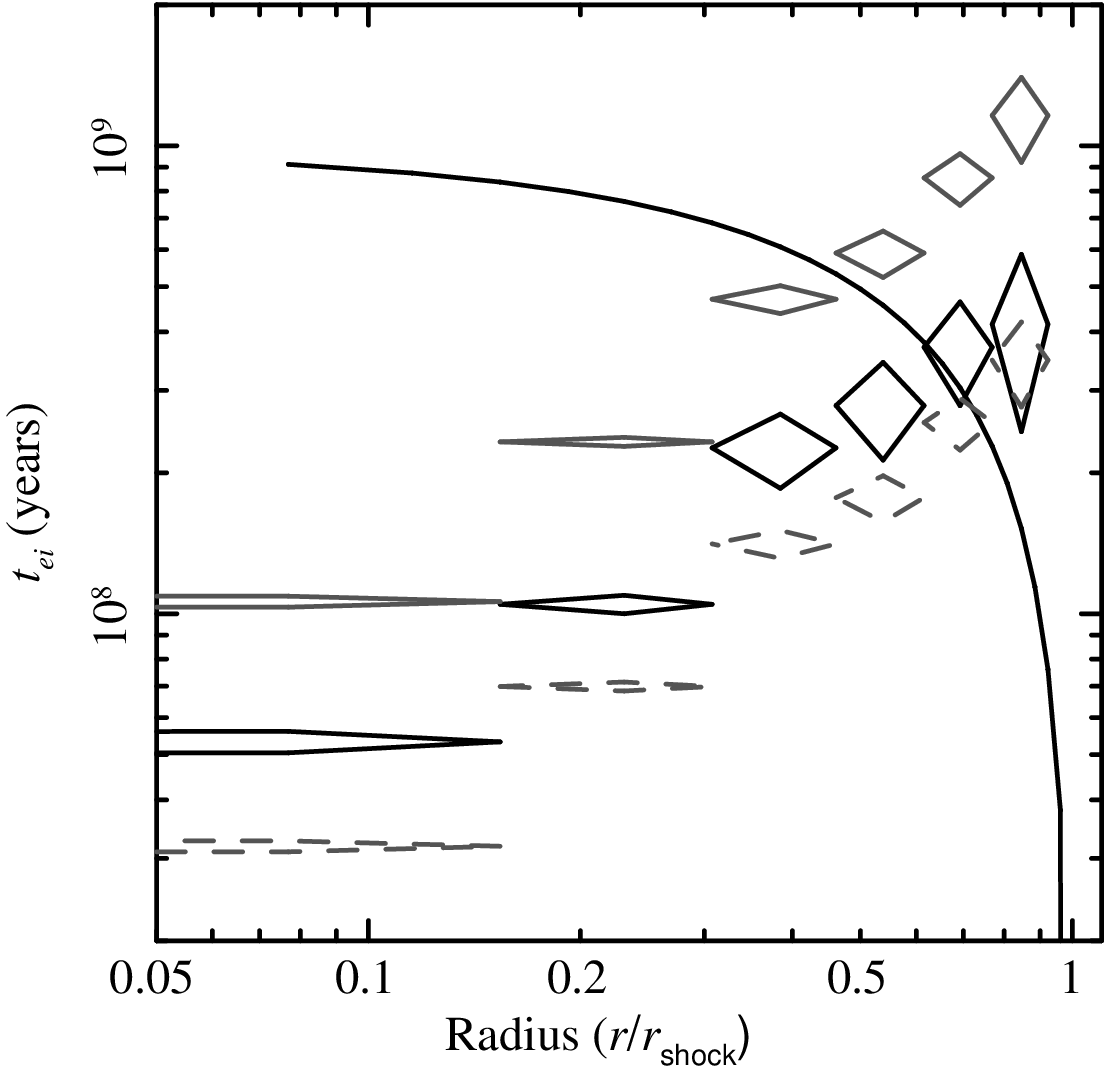}
\end{center}
\end{minipage}
\end{tabular}
\caption{
Radial profiles of pressure, entropy, entropy ratio to
 theoretical value, and equilibration time after a shock passage.  (a)
 (b) Black and gray diamonds show results of Suzaku and XMM-Newton,
 respectively.  Dashed vertical lines show approximate radial
 boundaries of the northwest radio relic.  In the entropy profile (b),
 the black solid line shows $K \propto r^{1.1}$ given by~\cite{voit03}.
 (c) Ratio of the measured entropy with reflrect to the theoretical value ($K \propto r^{1.1}$). 
 The horizontal scale is normalized by the virial radius.
 Black diamonds show Suzaku results for A3667, gray diamonds show A2142 results~\citep{akamatsu11}, 
 and 
 gray dashed diamonds 
 are A1413  values~\citep{hoshino10}, respectively. 
  (d) Ion-electron  equilibration time scale. 
 Horizontal scale is normalized by the radius of the radio relic.  
 The solid curve shows the time after a shock heating assuming a constant shock speed $v=1400$ km s$^{-1}$.  
 Black diamonds show the electron-ion equilibration time calculated by equation (3).  
 Gray solid and dashed diamonds show the ionization time for $n_{e}t=3\times10^{12} \rm and 1\times10^{13}$ s cm$^{-3}$, respectively. 
}
\label{fig:PE}
\end{figure*}

The present Suzaku data show steep jumps in temperature, surface brightness, and electron number density 
at the region of the radio relic.  
To verify the presence and properties of a shock, we estimate the ICM pressure.  
Fig~\ref{fig:PE}(b) shows the resultant radial profile of the gas pressure in A3667\@.  
Black and gray curves show the results from Suzaku and XMM-Newton, respectively.  
Since the temperature and the surface brightness both show a steep decline, 
with a somewhat less significant jump in the surface brightness, 
we see a clear pressure jump of a factor of $\gtrsim 3$.  
The spatial coincidence in the regions of the pressure drop and the radio relic
strongly suggests that there is a shock front in this region.

We estimate the Mach number based on the Suzaku data in the same way as \citet{finoguenov10}.  
The Mach number can be obtained by applying the Rankine-Hugoniot jump condition, assuming the ratio of
specific heats as $\gamma=5/3$, as
$\displaystyle\frac{1}{C}=\frac{3}{4{\cal M}^2}+\frac{1}{4}$, 
$\displaystyle\frac{T_2}{T_1}=\frac{5{\cal M}^4+14{\cal M}^2-3}{16{\cal M}^2}$
 and $\displaystyle\frac{P_{2}}{P_{1}}=1+\frac{2\gamma}{\gamma+1}({\cal M}^2-1)$.
$T_{1},T_{2}$ and $P_{1},P_{2}$  are the pre-shock and post-shock temperatures and pressure.
$C=n_{2}/n_{1}$ gives the shock compression.  
Using the observed jumps in temperature, electron density and pressure, 
we obtain $\cal M\rm=2.48\pm 0.46, 1.52\pm 0.10$ and $1.87\pm 0.27$, respectively.  
Table~\ref{tab:mach} compares our Mach numbers with the XMM-Newton values by \citet{finoguenov10}.
Suzaku results agree well with the previous XMM results of $\cal M\rm \sim 2.4$ with slightly smaller errors.

\begin{table}[t]
\begin{center}
\caption{Mach number estimates for the A3667 radio relic shock} \label{tab:mach}
\begin{tabular}{cccccccc}\hline \hline
 Mach number$^\ast$					& {\cite{finoguenov10}} 	& This work \\ \hline
 $\cal M$ (by $n_{e}$) 		&    1.68$ \pm$0.16 	&$1.52\pm 0.10$  	\\
 $\cal M$ (by $kT$) 	&	2.43$ \pm 0.77$	&$2.48\pm 0.46$   \\
 $\cal M$ (by Pressure)	& 	2.09$ \pm $0.47 	&1.87$\pm$0.27   \\ \hline
\multicolumn{3}{l}{$\ast$ We use the observed value as post shock region:$\timeform{25.2'}-\timeform{29.4'}$}\\
\multicolumn{3}{l}{and pre shock region: $\timeform{29.4'}-\timeform{33.6'}$.}
\end{tabular}
\end{center}
\end{table}

Merger shocks in clusters were reported in the Bullet cluster ($\cal
M\rm \sim 3.0$) and in A520 ($\cal M\rm \sim 2.1$).  
Although these Mach numbers and the one in A3667 are broadly similar,
their locations are very different as pointed out by \citet{finoguenov10}.  
In A3667, the shock is located very far from the cluster center at $\sim \timeform{30'}$ corresponding to 2 Mpc, 
in contrast to the shocks  in the Bullet cluster and A520\@.  
Using the gas density in the outer region ($n_e \sim 10^{-4}$ cm$^{-3}$), 
the pre-shock sound speed is $\sim v_{ss}=700$ km s$^{-1}$.  
Using the shock compression $C$ and this pre-shock velocity, we estimate a shock speed $v_{shock}=C
\cdot v_{ss}=1360\pm 120$ km s$^{-1}$, which is consistent with the previous 
XMM-Newton result ($v_{shock}=1210\pm 220$ km s$^{-1}$).
This shock speed is much lower than in other systems (Bullet: 4500 km s$^{-1}$, A520: 2300 km s$^{-1}$) 
and close to the speed of the cold front in the cluster central region ~$v_{cf}=1430\pm 290$ km~s$^{-1}$ \citep{vikhlinin01}.  
Recent hydrodynamic simulations of merging clusters predict that cluster merger events generate cold fronts 
in the central regions and 
arc-like shock waves 
toward the outskirts~\citep{mathis05, akahori10}.  
These predictions are in good agreement with our observational results.

Another possibility for the origin of the shock front is an
infall/accretion flow from large-scale structures.  
\citet{miniati01} predict strong accretion shocks with Mach numbers up to 1000.
Such shock waves caused by structure formation are also predicted in the Coma cluster \citep{ensslin98}.  
The Coma cluster shows a giant radio halo and a relic located far from the cluster center, and it exhibits a temperature jump 
across the radio edge \citep{brown11}.  
Although many simulations predict accretion shocks, 
the observational evidence is still very limited and further sensitive X-ray observations will be very important.

Based on these observational and theoretical arguments, 
the shock at the northwest radio relic of A3667 seems to be more in favor of being caused by a merger event.

\subsection{Entropy profile}
\label{sec:entropy}
The entropy of the ICM is used as an indicator of the energy acquired by
the gas.  The entropy is generated during the hierarchical assembly process.  
Numerical simulations indicate that self-similar growth of clusters commonly shows 
entropy profiles approximated by $r^{1.1}$ up to $r_{200}$~\citep{voit03}.  
Since entropy preserves a record of both the accretion history and the influence of non-gravitational processes
on the properties of its ICM, entropy is a good indicator of the cluster growth.  
Recent XMM \citep{pratt10} and Chandra results on the entropy profile showed the entropy slope within $r_{500}$, 
which is approximately $0.5~r_{200}$, to be in agreement with the predictions from numerical simulations.  
Recently, Suzaku has extended the entropy measurements close to $r_{200}$ for several clusters, and showed a
flattening or even a decrease at $r\gtrsim 0.5~r_{200}$~\citep{hoshino10, kawaharada10,akamatsu11,simionescu11}.

Comparing with relaxed clusters, we looked into the entropy profile of A3667\@.  
The entropy of the ICM is given by $K= kT n_e^{-2/3}.$ 
The entropy profile is shown, along with the XMM-Newton result (gray), in Fig.~\ref{fig:PE}.
The overplotted black line shows $K\propto r^{1.1}$.  The observed
entropy slope is consistent with the theoretical value between $\timeform{0'}- \timeform{10'}$.  
The slope becomes flatter at $r > \timeform{20'}$ and suddenly drops around the radio relic region.
This drop suggests the existence of non-thermal heating around the relic.

To compare the entropy profiles with the simulated slope of 1.1 and with other clusters, 
we calculated the ratio of the simulated and observed profiles.
The resultant profile is shown in Fig.~\ref{fig:PE}(c) together with
the ones for an other merger cluster (A2142:gray) and a relaxed cluster (A1413:dashed).  
There is a clear deviation from the simulation in the range $r > 0.4 r_{200}$.  
Such a feature has been also reported in other relaxed clusters such as A1795, PKS0745, A1689 \citep{george08,bautz09, kawaharada10}.  
The deviation from the simulated entropy curve in the cluster outskirts is apparently common to
many clusters including merging and non-merging systems.
Some physical conditions to explain the origin of the deviation were proposed~\citep{simionescu11, hoshino10}.
However, the interpretation of the observed features is not settled yet.

\subsection{The possibility of non-equilibrium ionization.}
Theoretical studies about a non-equilibrium state and an electron-ion
two-temperature structure of the ICM in merging galaxy clusters have been
carried out \citep{takizawa05,rudd09,akahori10}.  
Recent Suzaku results suggest non-equilibrium features characterized by a
different electron and ion temperature in the cluster outskirts \citep{wong09, hoshino10}.  
In merger clusters, shock heating and compression raise the X-ray luminosity in the outer regions, 
giving us a good opportunity to look into the non-equilibrium features and gas heating processes.

We examine the possibility of a deviation between ion and electron
temperatures.  In A3667, the existence of a shock is already suggested
in the radio relic area~\citep{finoguenov10}.  Assuming that the shock
propagates from the cluster center to the outer region, the region
just behind the shock may have a higher ion temperature than the electron one,
because it had not enough time to equilibrate.

According to \cite{finoguenov10}, we calculate electron-ion
equilibration time and ionization timescales.
The electron-ion equilibration time is estimated from the electron density and
temperature of ICM.  The ionization timescales can be estimated from
relative line intensities.  Because the lines are very weak in the
cluster outer region, we adopt the timescale for equilibrium
ionization as $n_{e}t=3\times10^{12} - 1\times10^{13}$ s cm$^{-3}$
given by \cite{fujita08} indicating that the outer region attains ionization equilibrium.

In fig~\ref{fig:PE}(d), solid diamonds show the electron-ion
equilibration times of A3667 based on the observed temperature and
density profiles.  
The gray and dashed diamonds in fig~\ref{fig:PE}(d) show the resultant values of  ionization timescales  for $n_{e}t=3\times10^{12}$ s cm$^{3}$ and
$1\times10^{13}$ s cm$^{3}$, respectively.
The horizontal axis of fig~\ref{fig:PE}(d) is normalized by the radio relic radius.  
The solid curve shows the expected time after the shock heating assuming a
constant shock speed $v_{shock}=1400$ km s$^{-1}$. 
The relaxation and ionization timescales, considering electron-ion Coulomb
collisions, are both longer than the expected elapsed time after the
shock heating around $r> 0.7 r_{\rm shock}$. This suggests that the
region near the radio relic would not have reached ion-electron
equilibrium. This can be confirmed with future spectroscopic
observations which will be able to determine ion temperature through
line width measurements.

\section{Summary}
We observed Abell~3667 with Suzaku XIS and detected the ICM
emission near the virial radius $r_{200}$ (2.3 Mpc $\sim 34'$).  
We confirm the jumps in temperature and surface brightness across the
northwest radio relic region, which is located far from the cluster center~(2 Mpc).  
Using the Rankine-Hugoniot jump condition and observed ICM pressure value, 
we evaluated the Mach number as ${\cal M}=1.87\pm 0.27 $, 
 consistent with the previous XMM-Newton result~\citep{finoguenov10}. 
The main results on A3667 are as follows;
\begin{itemize}
\item The ICM temperature gradually decreases toward the outer region
 from about 7 keV to 5 keV and then shows a jump to 2 keV at the radio relic.  
 The temperature profile inside the relic shows an
 excess over the mean values for other clusters observed with XMM-Newton.
\item There is a distinct structure of the emission measure
 distribution around the radio relic, characterized by two peaks
 around 0.9 and 4.0 keV\@.
\item The electron density profile shows a good agreement with the
 $\beta$-model ($\beta=0.54$) and also shows a significant jump
 across the radio relic.
\item The significant jump in temperature and density profile across
 the radio relic support the presence of a shock.  The estimated Mach
 number of the shock is ${\cal M}=1.87\pm 0.27 $ based on the
 pressure jump.
\item The entropy profile within about $0.5 r_{200}$ follows
 $r^{1.1}$, predicted by the accretion shock heating model.  The
 profile shows a significant jump around the radio relic, 
  suggesting the region is not thermally relaxed.
\item Based on the relaxation and ionization time needed after a shock heating, 
we discuss the possibility of $T_i$ being higher than $T_e$ around the radio relic region.
Because of the different parameter dependence of electron-ion equilibration time and non-equilibrium ionization,
there is the possibility that not non-equilibrium ionization state but $T_i$ and  $T_e$  are different.
\end{itemize}

These results show that A3667 is a dramatic merging cluster, with
the outskirts still  in the process of reaching thermal equilibrium
and the electrons in the northwest radio relic  being strongly heated by their merger shock.
Because of the existence of a large cold front and presence of the radio relic, 
A3667 is a promising target for future X-ray studies such as with ASTRO-H\@.

\bigskip The authors thank all the Suzaku team members for their support of the Suzaku project.  
We also thank an anonymous referee for a lot of constructive comments.
H. A. is supported by a Grant-in-Aid for Japan Society for the Promotion of Science (JSPS) Fellows
(22$\cdot$1582) and the JAXA ASTRO-H science team.  
H. A. deeply acknowledges hospitality in the SRON HEA group.

\appendix
\section{Solar Wind Charge eXchange}
\begin{figure*}[t]
(a)
\begin{center}
\includegraphics[scale=0.35]{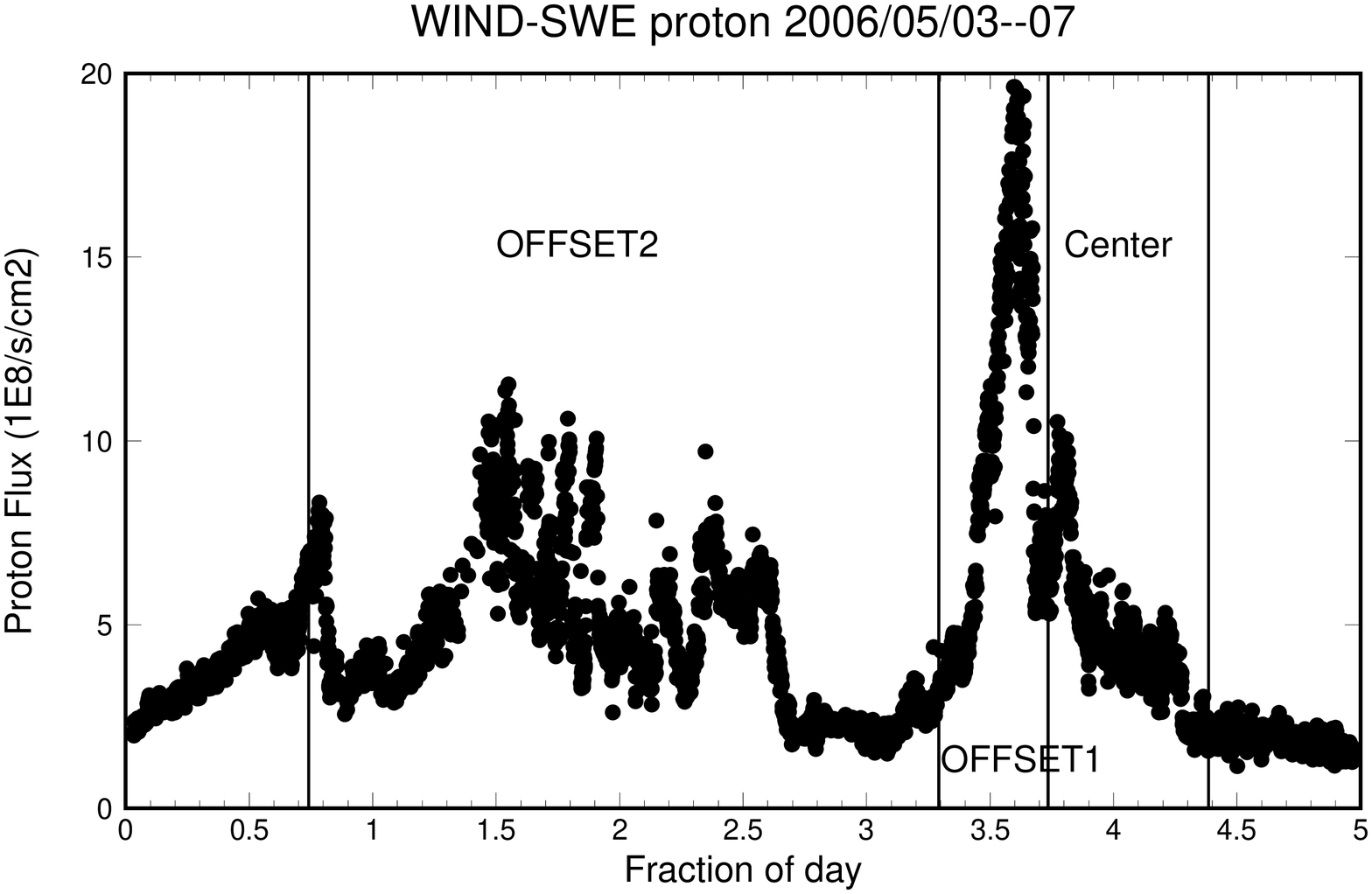}
\label{fig:proton}
\end{center}
\begin{tabular}{ccc}
\begin{minipage}{0.33\hsize}
(b) OFFSET2
\begin{center}
\includegraphics[scale=0.225,angle=-90]{off2xi1cor28_0.5_2.0lc_cor8.ps}
\end{center}
\end{minipage}
\begin{minipage}{0.33\hsize}
(c) OFFSET1
\begin{center}
\includegraphics[scale=0.225,angle=-90]{off1xi1cor28_0.5_2.0lc_cor8.ps}
\end{center}
\end{minipage}
\begin{minipage}{0.33\hsize}
(d) CENTER
\begin{center}
\includegraphics[scale=0.225,angle=-90]{centerxi1cor28_0.5_2.0lc_cor8.ps}
\end{center}
\end{minipage}
\end{tabular}
\caption{\footnotesize (a) Proton flux (speed times density) measured by the WIND-SWE when Suzaku observed A3667. 
 (b)-(d) Light curve of the Suzaku XIS1 0.5 - 2.0 keV band during each observations (OFFSET2, OFFSET1 and CENTER).}
\label{fig:lc}
\end{figure*}

To check the contamination effects of Solar Wind Charge eXchange(SWCX), 
we check the solar-wind proton flux during our Suzaku Observation~Fig. \ref{fig:lc}.
These plots are created by multiplying the proton speed and proton density of 
public WIND-SWE data (http://web.mit.edu/space/www/wind.html).
In the observation periods, proton flux has many flares.
Because the ICM emission is very strong in the center region of the cluster,
we can ignore effects of SWCX. 
In the cluster outskirts, ICM emission is very weak,  effects of SWCX is important.
In this analysis, the light curve of the most outer region (OFFSET 2) does not show 
time variation, therefore we assume effect of SWCX was very small.



\end{document}